\documentclass[fleqn,usenatbib]{mnras}

\usepackage{newtxtext,newtxmath}

\usepackage[T1]{fontenc}

\DeclareRobustCommand{\VAN}[3]{#2}
\let\VANthebibliography\thebibliography
\def\thebibliography{\DeclareRobustCommand{\VAN}[3]{##3}\VANthebibliography}

\usepackage{graphicx}	
\usepackage{amsmath}	

\title[Asteroseismology of RR~Lyrae stars]{Asteroseismology of RR~Lyrae stars with non-radial modes}

\author[H. Netzel et al.]{
Henryka Netzel,$^{1,2,3}$\thanks{E-mail: henia@netzel.pl}
Rados\l aw Smolec,$^{3}$
\\
$^{1}$Konkoly Observatory, Research Centre for Astronomy and Earth Sciences, E\"otv\"os Lor\'and Research Network (ELKH), MTA Centre of Excellence,	\\ H-1121 Konkoly Thege Mikl\'os \'ut 15-17, Budapest, Hungary\\
$^{2}$MTA CSFK Lend\"ulet Near-Field Cosmology Research Group\\ H-1121 Konkoly Thege Mikl\'os \'ut 15-17, Budapest, Hungary\\
$^{3}$Nicolaus Copernicus Astronomical Centre, Polish Academy of Sciences, Bartycka 18, 00-716 Warsaw, Poland
}

\date{Accepted XXX. Received YYY; in original form ZZZ}

\pubyear{2022}

\begin{document}
\label{firstpage}
\pagerange{\pageref{firstpage}--\pageref{lastpage}}
\maketitle

\begin{abstract}
The additional signals observed in the frequency spectra of the first-overtone RR~Lyrae stars, that form a period ratio around 0.61 with the period of the first overtone, are a common phenomenon for RRc and RRd stars, as well as for first-overtone classical Cepheids. The recently proposed model explains these signals as harmonics of non-radial modes of degrees 8 or 9 in the case of RR~Lyrae stars and 7, 8, or 9 in the case of classical Cepheids. We selected at least triple-mode RR~Lyrae stars pulsating in radial and non-radial modes for asteroseismic modeling. We assume the identification of the non-radial modes as predicted by the model. We calculated a dense grid of models for RR~Lyrae stars using envelope pulsation code. By matching first overtone period and period ratios we obtained physical parameters for the selected sample of triple-mode stars. It is the very first attempt of modeling RR~Lyrae stars with non-radial modes. We compared our results with predictions of stellar evolution theory, which resulted in a mass discrepancy more noticeable for long-period stars: pulsation masses seem higher than evolutionary masses. We compared metallicity estimates for RRc stars from modeled sample with metallicities determined spectroscopically for a sample of RRc stars in the solar neighbourhood: both distributions are consistent. 
\end{abstract}




\section{Introduction}

RR~Lyrae stars are low-mass population II stars from the classical instability strip. Typically, RR~Lyrae stars are single or double-mode pulsators of periods of $0.2-1$\,d and amplitudes up to 1\,mag in the $I$ band. The majority of them pulsate either in the radial fundamental mode (RRab type) or in the first overtone (RRc type). Double-mode pulsations in the fundamental and first-overtone radial modes are also known (RRd type). Significantly less common are double-mode pulsations in radial fundamental and second overtone \citep[see e.g.][]{moskalik2013,benko2010} or triple-mode pulsations in radial modes \citep{jurcsik2015}. \cite{molnar2012} reported simultaneous detection of fundamental mode, first overtone and ninth overtone in the RR Lyr, identification of which is supported by the non-linear models.

Apart from pulsation in radial modes, we observe additional phenomena in RR~Lyrae stars, e.g. the Blazhko effect or additional periodicities that do not correspond to radial pulsations. 

The Blazhko effect is a quasi-periodic modulation of amplitude and/or phase \citep{blazhko}. The incidence rate of the Blazhko effect among RRab stars is higher \citep[up to 50 per cent,][]{jurcsik2017,benko2010} than among RRc stars \citep[below 10 per cent,][]{netzel_blazhko}. The Blazhko effect was also observed in the double-mode stars \citep[anomalous RRd, e.g.,][]{soszynski.smolec2016}. 

Many new groups of multi-periodic RR~Lyrae stars were detected over the last few years thanks to both ground and space-based photometry. The additional signals detected in these stars do not correspond to radial pulsations and for the majority of them, the nature remains unknown. For the recent development on the multi-mode RR~Lyrae stars see \cite{smolec2017_pet_rev}.

One group of multi-mode RR~Lyrae stars is particularly interesting, the so-called RR$_{0.61}$ stars. RR$_{0.61}$ group consists of RRc or RRd stars. The period of the additional low-amplitude signal present in these stars forms a period ratio with the first-overtone period in a 0.60 -- 0.65 range. More than a thousand RR$_{0.61}$ stars are known up to date \citep[see e.g.][and references therein]{smolec2017_pet_rev,netzel_census,molnar2021}. Within the RR$_{0.61}$ group, the three well-resolved and nearly parallel sequences are observed in the Petersen diagram \citep[see fig. 3 in][]{netzel_census}. The explanation of the nature of signals in RR$_{0.61}$ stars was proposed by \cite{dziembowski2016}. In the proposed model, the signals forming the characteristic period ratio 0.60 -- 0.65 are harmonics of the non-radial f-modes trapped in the outer layers, which have degrees of  $\ell=8$ or 9. We note, that similar signals that form a period ratio with the first-overtone mode from a range of 0.60 -- 0.65 are also known among classical Cepheids \citep[e.g.,][]{soszynski2010,smolec.sniegowska2016,rajeev} and as proposed by \cite{dziembowski2016} they correspond to the harmonics of non-radial modes of degrees $\ell=7$, 8 or 9. Interestingly, in the anomalous Cepheid, XZ Cet, \cite{plachy2021} also detected two additional signals which form period ratios corresponding to the non-radial modes of degrees 8 and 9, according to the above explanation \citep[see fig. 6 in][]{plachy2021}. These modes are strongly trapped unstable modes (STU) and possibility of their excitation was discussed by \cite{vanhoolst1998}. Their driving rates are comparable with those of radial modes. Even though there is still no direct mode identification for these non-radial modes, the results based on photometric observations seem to be consistent with the predictions of the Dziembowski's model \citep[see][]{smolec.sniegowska2016,netzel_census}.

Even though asteroseismology plays an important role in studying other kinds of pulsating stars, so far, only a handful of RR~Lyrae stars and classical Cepheids have their physical parameters determined based on asteroseismic modeling. \cite{moskalik2005} performed asteroseismic modeling of radial triple-mode classical Cepheids.  \cite{molnar2015} used non-linear hydrodynamic pulsation models to obtain physical parameters for two RRd stars observed by the K2 space mission. \cite{molnar2012} discussed the prospects of non-linear asteroseismology of RR Lyrae stars, in particular of those showing period doubling effect. Assuming that the scenario proposed by \cite{dziembowski2016} is correct, we have RR~Lyrae stars that pulsate in radial and non-radial modes, identification of which is known. Moreover, some of them are triple-periodic. This, in turn, provides the very first opportunity to use asteroseismic modeling to study RR~Lyrae stars with non-radial modes. Additionally, the attempt at modeling of stars assuming the identification proposed by \cite{dziembowski2016} is yet another test of this theory. For a correct model, we expect to be able to match the periods of most stars and we expect reasonable values of the physical parameters, in the range expected for RR~Lyrae stars.

Asteroseismic modeling of RR~Lyrae stars results in estimation of their physical parameters. Particularly important is mass determination. Up to date, no RR~Lyrae star is known to be a member of an eclipsing binary system, which would provide an opportunity to measure the dynamical mass. The most promising candidate for an RR~Lyrae stars in a binary system turned out to be an object formed via a very different evolutionary channel and having a mass of 0.26~M$_\odot$, which is significantly smaller than expected for a canonical RR~Lyrae star \citep{pietrzynski2012}. The efforts to discover RR~Lyrae star in binary systems are ongoing and already resulted in a detection of a few candidates based on the light-travel time effect \citep{hajdu2021} or the analysis of proper motions \citep{kervella2019a,kervella2019b}. Still, however, we lack a direct mass determination for any RR~Lyrae star.  

In this paper, we present the results of the modeling of the triple-mode  RR$_{0.61}$ stars. We assumed the mode identification as predicted by \cite{dziembowski2016}. Selection of the initial sample of stars for modeling and calculation of theoretical models are described in Sec.~\ref{Sec.method}. Results of the modeling are presented in Sec.~\ref{Sec.results} and discussed in Sec.~\ref{Sec.discussion}. Sec.~\ref{Sec.conclusions} concludes the work.

\section{Method}\label{Sec.method}

RR~Lyrae stars are evolved giants. Their internal structure differs significantly from the main-sequence stars. In particular, they have very concentrated interiors and extended outer layers. As was shown by \cite{epstein1950}, pulsations of giant stars have relatively small amplitudes in interiors. Mostly the outer layers impact the pulsation properties. \cite{dziembowski1977} showed that non-radial modes of frequencies close to the frequency of the radial fundamental mode or the radial first overtone, can be studied considering only the outer layers of the star. In other words, only the structure of the envelope can be used in the pulsation code.

\cite{paczynski1969} provided the structure equations for an envelope with the assumption of a gray atmosphere and convection described by the mixing length theory. Resulting envelope structure is then used in the Warsaw pulsation code by \cite{dziembowski1977}. In the pulsation equations, frozen-in approximation for perturbation of convective flux is used.

\subsection{Studied sample}

For the asteroseismic modeling, we chose RR~Lyrae stars pulsating in at least three modes. These are RRc stars with two additional non-radial modes or RRd stars with at least one additional non-radial mode. In Table~\ref{tab.rr_kand} we present RRc stars with two additional signals corresponding to the top and the bottom sequence on the Petersen diagram. These stars were selected from the sample analyzed by \cite{netzel_census}. An additional criterion for the selection of these stars was that the signal-to-noise ratio of the additional signals is above 5. This criterion was chosen to avoid the possible, but relatively low risk of modeling a false positive detection. For the modeling, we consider only period ratios that are formed by the harmonics of non-radial modes. Therefore, it was not important whether the non-radial modes themselves were detected in these stars. As pointed out by \cite{dziembowski2016}, the non-radial modes of degrees 8 and 9 are characterized by stronger variability than their harmonics. Therefore, it is more difficult to estimate the representative frequency of the non-radial mode that would have to be reproduced in the modeling. We refer the reader to \cite{netzel_census} for the discussion of sometimes complex appearance of signals related to the excitation of the discussed non-radial  modes in the frequency spectrum. In total, based on the analysis from \cite{netzel_census}, we selected 32 RRc stars with two additional signals corresponding to the harmonics of the non-radial modes. Table~\ref{tab.rr_kand} provides IDs of these stars, periods of the first overtone, both period ratios of harmonics of the non-radial modes with the period of the first overtone defined as:

\begin{equation}
    R_\ell = \frac{0.5P_\ell}{P_{\rm 1O}}.
    \label{eq.R}
\end{equation}
We also provided values of signal-to-noise ratio for the signals corresponding to the harmonics of non-radial modes.

We analyzed light curves of RRc stars from \cite{plachy2019} based on the K2 data for sectors 3, 4, 5, and 6 to select additional stars for the modeling. We selected 11 more RRc stars with two additional signals corresponding to the harmonics of the non-radial modes.

Two RRc stars with both additional signals were found and analyzed by \cite{smolec2017} based on the ground-based photometry for the globular cluster NGC~6362. These stars are also included in the analysis. Information on periods and period ratios for the RRc stars from NGC~6362 cluster and the K2 photometry are presented in Table~\ref{tab.rr_kand2}.

We selected RRd stars with at least one harmonic of the non-radial mode detected in the frequency spectrum. Basic information on these stars are presented in Table~\ref{tab.rrd_kand}. Eleven RRd stars were selected based on the analysis by \cite{netzel_census}. In 10 stars there is one additional signal. In five stars, the additional signal corresponds to the harmonic of the non-radial mode of degree 8 and in five stars the signal corresponds to the harmonic of the non-radial mode of degree 9. In one star, OGLE-BLG-RRLYR-10769, there are two additional signals corresponding to the harmonics of both non-radial modes. We selected additional RRd stars with non-radial modes analyzed in the literature. We selected three stars from the results obtained by \cite{jurcsik2015} for globular cluster M3. Two of them have signals corresponding to the non-radial mode of degree 9. The third star is quadruple-mode and pulsates in the fundamental mode, first overtone, second overtone, and has a harmonic of the non-radial mode of degree 9. Three more stars were found in the analysis of the space-based photometry. One RRd star was detected by \cite{chadid2012} based on the CoRoT data, and two more RRd stars were detected in the analysis of the {\it Kepler} data by \cite{molnar2015}. We included also the first known RR~Lyrae star showing this additional signal, AQ Leo \citep{gruberbauer2007}.

\begin{table}
 \centering
  \caption{RRc stars with two additional signals corresponding to the harmonics of the non-radial modes. Consecutive columns provide star's ID (OGLE-BLG-RRLYR-), period of the first overtone, and period ratios for both observed signals (frequencies of the signals are adopted from the Gaussian function fit -- see text for more details), and signal-to-noise ratios for both additional signals.}~\\
  \label{tab.rr_kand}
  \centering
  \begin{tabular}{@{}rlllrr@{}}

  \hline
ID & $P_{\rm 1O}$ [d] & $R_8$ & $R_9$ & SNR$_8$ & SNR$_9$  \\ 
\hline
5202	&	0.31386605(5)	&	0.63564	&	0.61272	&	5.7	&	5.3	\\
6352	&	0.3162666(2)	&	0.63122	&	0.61319	&	8.2	&	12.8	\\
6802	&	0.31830253(5)	&	0.63138	&	0.61264	&	5.0	&	7.3	\\
6922	&	0.30877929(3)	&	0.63080	&	0.61256	&	5.2	&	6.7	\\
7047	&	0.30496436(2)	&	0.63019	&	0.61229	&	5.1	&	5.5	\\
7665	&	0.26829993(4)	&	0.63135	&	0.61897	&	5.2	&	9.1	\\
7803	&	0.31390033(3)	&	0.63113	&	0.61235	&	5.7	&	14.7	\\
7806	&	0.31899507(3)	&	0.63125	&	0.61310	&	6.4	&	7.2	\\
8002	&	0.30879120(3)	&	0.63060	&	0.61264	&	5.4	&	8.0	\\
8475	&	0.31094850(5)	&	0.63113	&	0.61316	&	7.5	&	9.9	\\
8799	&	0.31734930(3)	&	0.63105	&	0.61296	&	5.5	&	8.2	\\
8826	&	0.31251553(3)	&	0.63091	&	0.61202	&	7.5	&	7.4	\\
8920	&	0.31500873(6)	&	0.63109	&	0.61316	&	5.2	&	6.9	\\
8980	&	0.31356147(3)	&	0.63124	&	0.61259	&	5.2	&	6.8	\\
9444	&	0.32376783(4)	&	0.63123	&	0.61332	&	6.7	&	5.0	\\
9733	&	0.31378022(3)	&	0.63130	&	0.61239	&	6.1	&	8.2	\\
10119	&	0.31540045(3)	&	0.63111	&	0.61304	&	6.1	&	5.3	\\
10262	&	0.31622131(1)	&	0.63114	&	0.61309	&	6.9	&	6.4	\\
10534	&	0.31529923(3)	&	0.63154	&	0.61325	&	7.5	&	8.6	\\
10614	&	0.31448656(4)	&	0.63107	&	0.61268	&	5.3	&	6.3	\\
11072	&	0.31601750(4)	&	0.63119	&	0.61306	&	5.7	&	5.4	\\
11547	&	0.30619293(2)	&	0.63067	&	0.61146	&	5.4	&	12.3	\\
11621	&	0.31723543(4)	&	0.63115	&	0.61308	&	5.5	&	5.8	\\
11728	&	0.31518348(3)	&	0.63122	&	0.61329	&	6.0	&	7.0	\\
11913	&	0.31835441(3)	&	0.63121	&	0.61299	&	9.1	&	9.4	\\
12261	&	0.30506798(2)	&	0.63101	&	0.61309	&	6.8	&	18.1	\\
12769	&	0.31341085(2)	&	0.63128	&	0.61319	&	6.5	&	13.6	\\
12776	&	0.31535057(2)	&	0.63100	&	0.61317	&	5.1	&	7.3	\\
12972	&	0.32094619(3)	&	0.63083	&	0.61254	&	6.1	&	8.1	\\
13156	&	0.31633834(3)	&	0.63129	&	0.61291	&	6.0	&	9.3	\\
31736	&	0.30524110(3)	&	0.63093	&	0.61297	&	7.0	&	7.4	\\
32213	&	0.24889440(5)	&	0.63232	&	0.61275	&	5.9	&	5.9	\\
\hline
\end{tabular}
\end{table}

\begin{table}
 \centering
  \caption{Additional RRc stars with two additional signals corresponding to the harmonics of the non-radial modes. Two stars come from the globular cluster NGC~6362 \protect\cite{smolec2017}. Eleven stars were selected by the authors based on the K2 photometry prepared by \protect\cite{plachy2019}. Consecutive columns provide the star's ID, period of the first overtone, and period ratios for both observed signals. First overtone periods for NGC~6362 variables are given to the last significant digit.}~\\
  \label{tab.rr_kand2}
  \centering
  \begin{tabular}{@{}llll@{}}

  \hline
ID & $P_{\rm 1O}$ [d] & $R_8$ & $R_9$  \\ 
\hline
NGC 6362 V17	&	0.31460473 	&	0.63089	&	0.61165	\\
NGC 6362 V33	&	0.30641758	&	0.63089	&	0.61300	\\
EPIC 210438688	&	0.3280081(9)	&	0.63115	&	0.61214	\\
EPIC 211701322	&	0.3391090(9)	&	0.63169	&	0.61400	\\
EPIC 211728918	&	0.3394932(6)	&	0.63211	&	0.61481	\\
EPIC 212316775	&	0.3304611(6)	&	0.63231	&	0.61420	\\
EPIC 212347262	&	0.3342333(8)	&	0.63195	&	0.61429	\\
EPIC 212352472	&	0.3359155(8)	&	0.63203	&	0.61557	\\
EPIC 212419731	&	0.3355586(7)	&	0.63181	&	0.61520	\\
EPIC 212448152	&	0.331029(1)	&	0.63167	&	0.61502	\\
EPIC 212613425	&	0.368725(1)	&	0.63101	&	0.61279	\\
EPIC 212684145	&	0.3524773(3)	&	0.63214	&	0.61513	\\
EPIC 212824246	&	0.313374(1)	&	0.63091	&	0.61315	\\
\hline
\end{tabular}
\end{table}

\begin{table*}
 \centering
  \caption{RRd stars with the additional signal corresponding to the harmonic of the non-radial modes. Consecutive columns provide the star's ID, periods of the fundamental mode and the first overtone, period ratios of the detected additional signals with the first overtone, and literature source for each star.}~\\
  \label{tab.rrd_kand}
  \centering
  \begin{tabular}{@{}llllll@{}}
  \hline
ID & $P_{\rm F}$ [d] & $P_{\rm 1O}$ [d] & $R_9$ & $R_8$ & Reference \\ 
\hline
OGLE-BLG-RRLYR-09258	&	0.3793278(1)	&	0.27766821(4)	&	-	&	0.63079	&	\cite{netzel_census}	\\
OGLE-BLG-RRLYR-10369	&	0.4348707(2)	&	0.32170650(5)	&	-	&	0.63179	&	\cite{netzel_census}	\\
OGLE-BLG-RRLYR-10744	&	0.478116(1)	&	0.35576412(8)	&	0.61417	&	-	&	\cite{netzel_census}	\\
OGLE-BLG-RRLYR-10796*	&	0.4313758(2)	&	0.31909935(4)	&	0.61243	&	0.63087	&	\cite{netzel_census}	\\
OGLE-BLG-RRLYR-11234	&	0.474381(4)	&	0.35300247(8)	&	0.61403	&	-	&	\cite{netzel_census}	\\
OGLE-BLG-RRLYR-11981	&	0.5901921(4)	&	0.43938460(5)	&	0.61656	&	-	&	\cite{netzel_census}		\\
OGLE-BLG-RRLYR-13198	&	0.434438(1)	&	0.32174412(5)	&	-	&	0.63158	&	\cite{netzel_census}	\\
OGLE-BLG-RRLYR-13666	&	0.4134416(2)	&	0.30495950(6)	&	-	&	0.63155	&	\cite{netzel_census}	\\
OGLE-BLG-RRLYR-13721	&	0.457128(1)	&	0.3393893(6)	&	-	&	0.63591	&	\cite{netzel_census}	\\
OGLE-BLG-RRLYR-14029	&	0.4299453(3)	&	0.31798614(6)	&	0.61275	&	-	&	\cite{netzel_census}	\\
OGLE-BLG-RRLYR-14031	&	0.576121(1)	&	0.4297653(2)	&	0.61847	&	-	&	\cite{netzel_census}	\\
CoRoT 0101368812	&	0.48804	&	0.36360	&	0.61409	&	-	&	\cite{chadid2012}	\\
EPIC 60018653	&	0.53943	&	0.40231	&	0.61634	&	-	&	\cite{molnar2015}	\\
EPIC 60018662	&	0.55900	&	0.41745	&	0.61701	&	-	&	\cite{molnar2015}	\\
M3 V13*	&	0.47950	&	0.35072	&	0.61370	&	-	&	\cite{jurcsik2015}		\\
M3 V68	&	0.47850	&	0.35597	&	0.61450	&	-	&	\cite{jurcsik2015}	\\
M3 V87	&	0.48017	&	0.35749	&	0.61770	&	-	&	\cite{jurcsik2015}	\\
AQ Leo &  	0.54976 &	0.41014 &	0.62105 &	-	&	\cite{gruberbauer2007}	\\
\hline
\multicolumn{6}{l}{* - stars with more than three independent signals in the frequency spectra, see text for details.}  \\ 
\hline
\end{tabular}
\end{table*}

\subsection{Grid of models}

The first step of the analysis was to calculate a grid of models used to fit the observed stars in the next step. The models were calculated using the Warsaw envelope pulsation code \citep{dziembowski1977}. Input parameters for the code are mass, luminosity, effective temperature, hydrogen abundance, $X$, and metal abundance, $Z$. We chose mass range of $0.5-0.9\,{\rm M}_\odot$ with a step $0.01\,{\rm M}_\odot$, luminosity range $\log L/{\rm L_\odot} \in \langle 1.3, 1.8 \rangle$ dex with a step of $0.01$\,dex, effective temperature range $\log T_{\rm eff} \in \langle 3.75, 3.91 \rangle$ dex with a step of $0.005$\,dex, and metallicities from ${\rm [Fe/H]}=-3.0$ dex to ${\rm [Fe/H]}=+0.1$ dex with a step of 0.05\,dex. Hydrogen and metal abundances were calculated based on the metallicities, ${\rm [Fe/H]}$, using:

\begin{equation}
    Z=10^{{\rm [Fe/H]}} Z_\odot
\end{equation}
and 
\begin{equation}
    Y=Y_{\rm p} + \frac{\Delta Y}{\Delta Z} Z,
\end{equation}
where we assumed primordial helium abundance $Y_{\rm p} = 0.2485$ \citep{cyburt2004}, and helium enrichment law $\Delta Y / \Delta Z = 1.5$ \citep{choi2016}. Solar metal abundance was set to $Z_\odot = 0.0134$ based on \cite{asplund2009}. We used OPAL opacities \citep{iglesias1996}. The mixing-length parameter, $\alpha_{\rm MLT}$ was set to 1.5. We neglected $\alpha$-enhancement. As stated by \cite{kovacs2021}, the effect of $\alpha$-enhancement on periods has an order of magnitude of $10^{-3}$ only, hence has little influence on the results of modeling.

From the selected ranges of input parameters, we chose only those combinations that put the model inside the theoretically predicted instability strip for RR~Lyrae stars. The envelope pulsation code used in this work uses the frozen-in convection treatment. Hence, it is possible to estimate the position of the blue edge of the instability strip, but it is not possible to obtain the position of the red edge. For this reason, we adopted the positions of blue and red edges of the instability strip as predicted theoretically by \cite{marconi2015}, who used the code taking into account convection-pulsation interactions \citep{bono_stellingwerf1994}. For a given metallicity, \cite{marconi2015} provide expressions for the position of the blue edge of the instability strip for the first overtone, and the position of the red edge for the fundamental mode. We selected only those sets of input parameters that are within the predicted boundaries.

\begin{figure}
    \centering
    \includegraphics[width=\columnwidth]{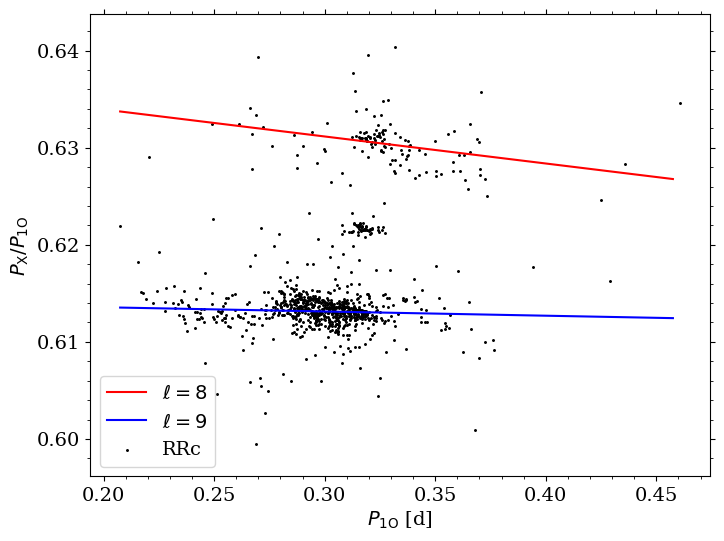}
    \caption{Fit to the period ratio -- first-overtone period relation for RR$_{0.61}$ analyzed by \protect\cite{netzel_census}. Fitted lines are described by Eq.~\ref{eq.fit8} and Eq.~\ref{eq.fit9}.}
    \label{fig:rrc_fit}
\end{figure}

The pulsation code requires an approximate value of the dimensionless frequency of the non-radial mode. To obtain this value, we first calculated the period of the radial first-overtone for each model. Then we used the value of the period of the first overtone to estimate the period of the non-radial mode. For this purpose, we used a fit to observed relation between period ratios, $R_8$ and $R_9$, and first overtone period, as presented in Fig.~\ref{fig:rrc_fit}. We got the following relations:

\begin{equation}
R_8=-0.02782713 P_{\rm 1O} [\rm d]+0.6395014,
\label{eq.fit8}
\end{equation}

\begin{equation}
R_9=-0.0043501924 P_{\rm 1O} [\rm d]+0.6144396.
\label{eq.fit9}
\end{equation}
Then, the calculated periods were transformed to dimensionless frequency by using the expression:
\begin{equation}
\sigma = \frac{1}{P} \sqrt{\frac{\uppi} {G \overline{\rho}}}, 
\end{equation}
where $\overline{\rho}$ is a mean stellar density, $P$ is a period, and $G$ is a gravitational constant. According to \cite{dziembowski2016}, non-radial modes observed in RR~Lyrae stars are strongly trapped in the envelope and have the highest driving rates. To choose the best-trapped pulsation mode for a given model, we have initiated pulsation calculations with initial non-radial mode frequency in the range $\langle \sigma-0.1, \sigma+0.1 \rangle$ with a step of $\Delta \sigma = 0.002$. This approach led to a few modes of the same degree and slightly different driving rates. For each model, we chose a mode with the highest driving rate.

The envelope code, that was used for calculations, returns periods, and growth rates for the fundamental mode, first overtone, and non-radial modes of degrees 8 and 9. For comparison with observations, we used the period ratios between the harmonic of the non-radial mode and the first overtone, for two reasons. Firstly, the harmonics of non-radial modes are more often detected. It means that the sample for modeling can be more numerous. Secondly, structures corresponding to the harmonics of non-radial modes in power spectra are narrower than structures corresponding to non-radial modes \citep[see e.g.][]{netzel_census}. As discussed by \cite{dziembowski2016} using the harmonics of non-radial modes for estimating the frequency is a more reliable method, as for the non-radial modes themselves the interactions between $2\ell+1$ components of the multiplet make the observed structures in power spectra even more complex \citep[see also][]{buchler.goupil1995}. In the linear pulsation analysis we also neglected non-linear interactions of the modes and we did not include rotation. We note, that non-linear periods differ from linear, but the difference is small, smaller than $10^{-4}$ \citep{smolecphd}. The difference between non-linear and linear period ratios of radial modes in RRd stars is also small \citep{kollath.buchler2001}.

\begin{figure}
    \centering
    \includegraphics[width=\columnwidth]{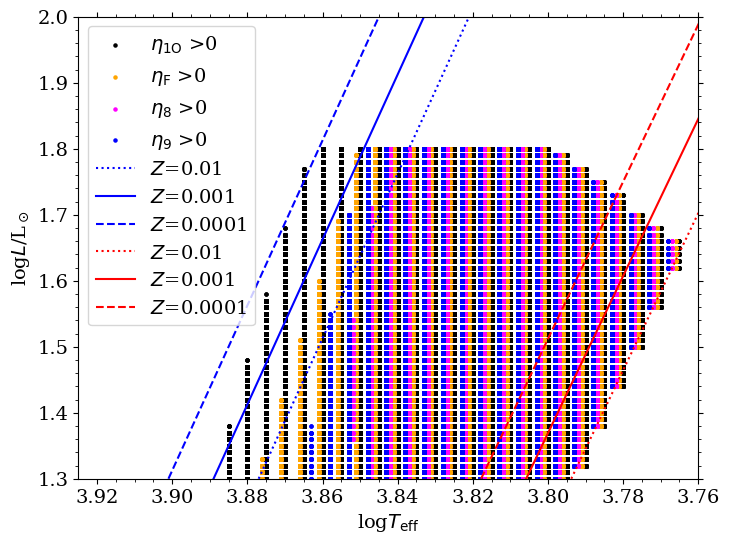}
    \caption{The calculated grid of theoretical pulsation models of RR~Lyrae stars with non-radial modes of degrees 8 and 9. Edges of the instability strip for three values of metallicity are based on \protect\cite{marconi2015}. Models with linearly unstable modes are plotted with different colors as indicated in the key. Each set of points corresponds to the same effective temperature as for the black point. The shift in effective temperature was used for better visualization.}
    \label{fig:grid_inst}
\end{figure}

The calculated grid consists of 2\,161\,303 models with different masses, metallicities, effective temperatures, and luminosities within the boundaries of the instability strip determined by \cite{marconi2015}. We rejected all models for which the period of the first overtone is longer than 0.5\,d because we do not have stars with periods that long in the modelled sample. This rejection resulted in the remaining 1\,919\,657 models. These models are presented in the HR diagram in Fig.~\ref{fig:grid_inst}. Models with certain modes linearly unstable are marked with different colors as indicated in the legend. Note, that points indicating different unstable modes in the models of the same physical parameters are slightly moved in effective temperature for better visualization.

For modeling of RRc stars, we chose only those models in which first overtone and non-radial modes of degrees 8 and 9 are unstable. This sample consists of 993\,900 models. For modeling of RRd stars, we chose two sets of models. The first set consists of models in which the fundamental mode, first overtone, and non-radial mode of degree 8 are unstable. In this set, we have 990\,690 models. The second set consists of models in which the fundamental mode, first overtone, and non-radial mode of degree 9 are unstable. There are 1\,402\,664 such models. One RRd star in the sample, OGLE-BLG-RRLYR-10796, shows harmonic of both non-radial modes. For modeling this star we chose a set of 990\,674 models. 

\begin{figure}
    \centering
    \includegraphics[width=\columnwidth]{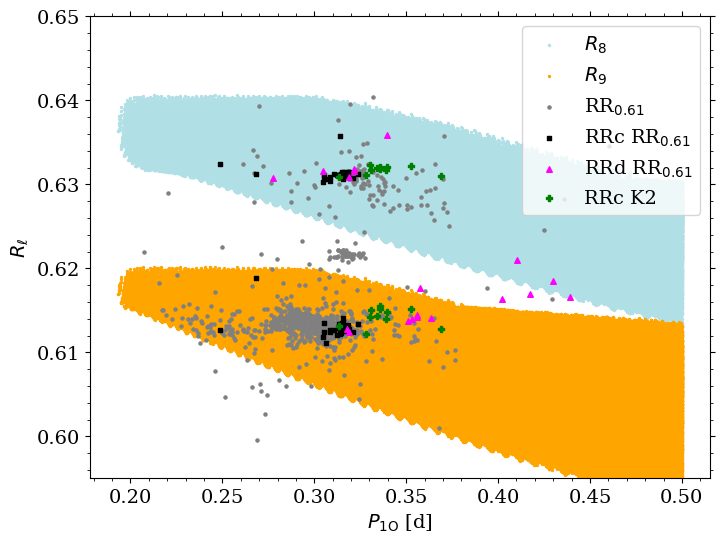}
    \caption{The grid of pulsation models presented in the form of the Petersen diagram. Blue points correspond to models with unstable non-radial mode of degree 8. Orange points correspond to models with unstable non-radial mode of degree 9. RR$_{0.61}$ stars observed towards the Galactic bulge are plotted with gray points \protect\citep{netzel_census}. Black squares, magenta triangles, and green pluses correspond to RRc and RRd stars selected for the modeling (Tables~\ref{tab.rr_kand}--\ref{tab.rrd_kand}).}
    \label{fig:grid_pet}
\end{figure}

In Fig.~\ref{fig:grid_pet} we present the grid of models in the form of a Petersen diagram together with RR$_{0.61}$ stars. Stars plotted with gray symbols were analyzed in \cite{netzel_census}. Stars selected from this sample for the present analysis are marked with black symbols. Additional RRc stars selected in the K2 data and presented in Table~\ref{tab.rr_kand2} are marked with green symbols. With magenta symbols, we plotted RRd stars presented in Table~\ref{tab.rrd_kand}. Models with unstable non-radial mode of degree 8 are plotted with blue points, and those with mode of degree 9 are plotted with  orange points. Based on the comparison of the position of RR$_{0.61}$ stars with the grid of models, we excluded four stars from further analysis. These stars are OGLE-BLG-RRLYR-11981, EPIC~60018653, EPIC~60018662, and V87 (from M3) and are located between the two sequences formed by the models. Hence, the sample for the modeling consists of 45 RRc stars and 14 RRd stars.

In Fig.~\ref{fig:grid_pet}, there are two RRd stars with the longest periods in the analysed sample. These are AQ~Leo and OGLE-BLG-RRLYR-14031. Period ratios for these stars are 0.6211 and 0.6185, respectively. Based on the criteria for the distinction between sequences formed by RR$_{0.61}$ stars, discussed by \cite{netzel_census}, AQ~Leo would belong to the middle sequence and OGLE-BLG-RRLYR-14031 to the bottom sequence. However, comparison with the models shows that these stars might belong to the top sequence. It means that the additional signals in these stars would correspond to the harmonic of the non-radial mode of degree 8. This classification was adopted in the modeling.

\subsection{Fitting of the models}

\begin{figure}
    \centering
    \includegraphics[width=\columnwidth]{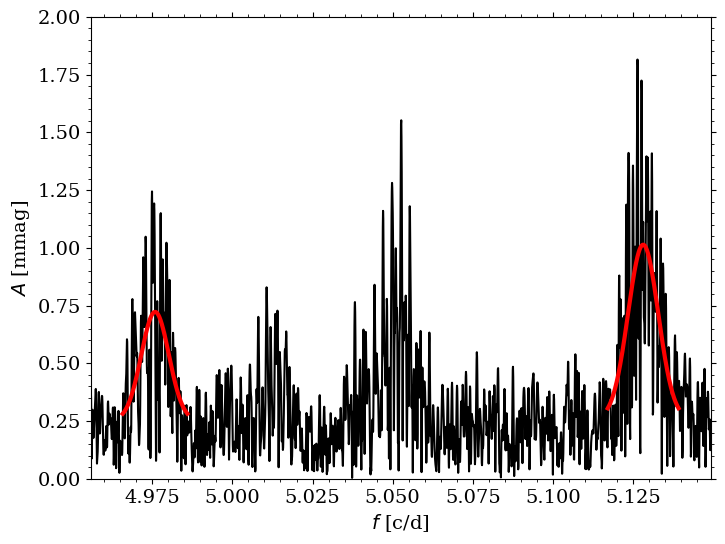}
    \caption{Example of the Gaussian fit (red lines) to the additional signals corresponding to the harmonics of the non-radial modes for OGLE-BLG-RRLYR-06802.}
    \label{fig:gauss}
\end{figure}

Due to the non-stationary nature of non-radial modes observed in RR$_{0.61}$ stars, in their frequency spectra we observe wide signals instead of single coherent peaks; both for the modes and for their harmonics \citep[][and references therein; Fig.~\ref{fig:gauss}]{netzel_census}. Hence, the estimation of the frequency of the harmonic is not straightforward and in some cases adopting the frequency of the highest peak in a group can be misleading. Another way to determine the representative value of the frequency is to fit the Gaussian function to the group of peaks and adopt its centroid as the frequency. We used this approach to estimate the frequency of harmonics of non-radial modes in the RRc stars from the sample prepared by \cite{netzel_census}. An example of the Gaussian fit to the harmonics of non-radial modes in an RRc star is presented in Fig.~\ref{fig:gauss} (in Table~\ref{tab.rr_kand}, values of period ratios based on such fits are given).

In the case of RRc stars from the K2 mission (Table~\ref{tab.rr_kand2}) we did not fit a Gaussian. Due to the short time of observations, around 80\,d, the wide structures connected to variability are not present in the spectra. For other analysed RRc stars, we used the values of periods and period ratios based on the literature (as given in Table~\ref{tab.rr_kand2}). RRd stars (Table~\ref{tab.rrd_kand}) were not fit with the Gaussian function as well.

To select the best model for each of the modeled stars, we chose the model minimizing the $D$ parameter defined as follows:

\begin{equation}
D^2=\left( \frac{P_{\rm F}^{\rm m} }{P_{\rm F}^{\rm o}} -1 \right)^2 + 
\left( \frac{P_{\rm 1O}^{\rm m}}{P_{\rm 1O}^{\rm o}} -1 \right)^2 +
\left( \frac{R_{\rm 8}^{\rm m}}{R_{\rm 8}^{\rm o}} -1 \right)^2 +
\left( \frac{R_{\rm 9}^{\rm m}}{R_{\rm 9}^{\rm o}} -1 \right)^2,
\label{Eq.chi2}
\end{equation}
where subscript `F' corresponds to fundamental mode, `1O' corresponds to the first overtone, `m' corresponds to theoretical values, and `o' to observed. $R_8$ and $R_9$ are period ratios defined as in Eq.~\ref{eq.R}. For each star, we used parts of the Eq.~\ref{Eq.chi2} relevant for a given pulsation type.

\section{Results}\label{Sec.results}
\begin{figure}
    \centering
    \includegraphics[width=\columnwidth]{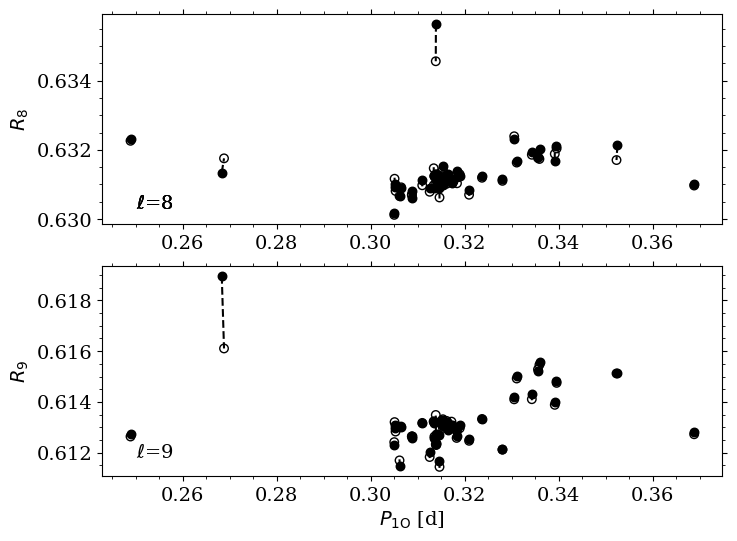}
    \caption{Comparison of the observed and calculated period and period ratios for RRc stars selected for the modeling. Two symbols are plotted for each star. Observed values are plotted with filled symbols, calculated values from the best fitting models are plotted with empty symbols. A dashed line is used for the stars for which fits are considered not satisfactory.}
    \label{fig:pet_comp_rrc}
\end{figure}

In the Petersen diagram in Fig.~\ref{fig:pet_comp_rrc} we present differences between observed values and theoretical values from the best-fitting models for RRc stars. Observed period ratios are marked with filled symbols, while best matching models with open symbols. We considered models to have good fits when $D^2<5 \cdot 10^{-7}$. Based on the comparison of fitted and observed values of periods and period ratios we excluded from the RRc sample three stars: OGLE-BLG-RRLYR-05202, OGLE-BLG-RRLYR-07665, and EPIC~212684145, for which $D^2>5 \cdot 10^{-7}$. In Fig.~\ref{fig:pet_comp_rrc}, corresponding points are connected with dashed lines. Possible reasons for lack of good fits for some stars are discussed in Sec.~\ref{Sec.discussion}.

\begin{figure}
    \centering
    \includegraphics[width=\columnwidth]{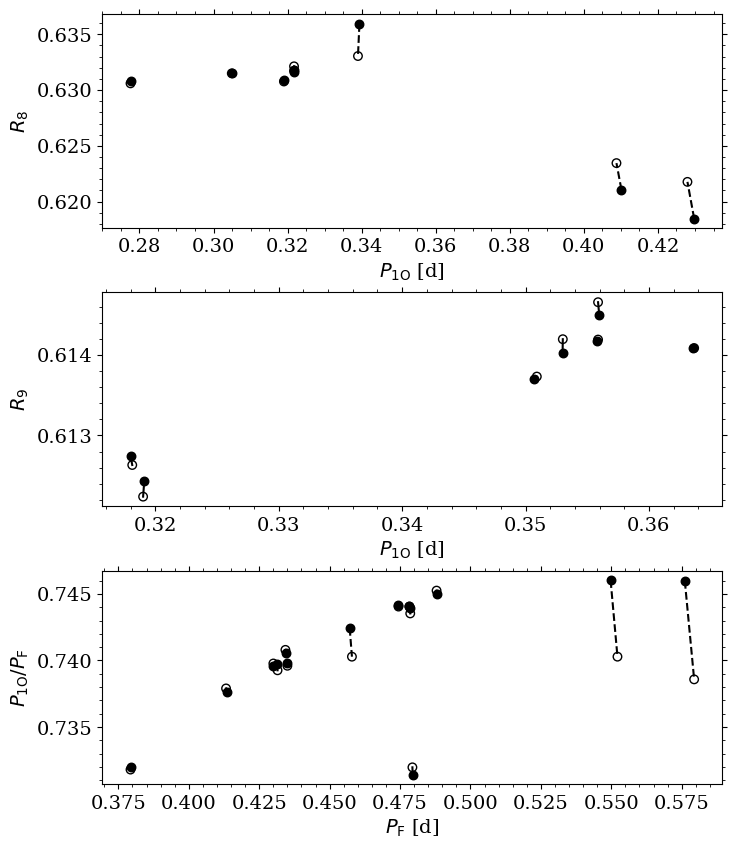}
    \caption{The same as Fig.~\ref{fig:pet_comp_rrc}, but for RRd stars.}
    \label{fig:pet_comp_rrd}
\end{figure}

In Petersen diagram in Fig.~\ref{fig:pet_comp_rrd} we present differences between the observed and fitted values of periods and period ratios for RRd stars. As for RRc stars, fits that are considered to be satisfactory when $D^2<5 \cdot 10^{-7}$. Based on this comparison we excluded three stars from the RRd sample: OGLE-BLG-RRLYR-13721, OGLE-BLG-RRLYR-14031, and AQ~Leo. This means that we were not able to obtain good fit for any RRd star with period of the first overtone longer than 0.4\,d.  The only quadruple-mode star, OGLE-BLG-RRLYR-10796, has a satisfactory fit corresponding to the parameter $D^2=2.2\cdot 10^{-7}$.

We were able to obtain satisfactory fits, and hence physical parameters for 42 RRc stars and 11 RRd stars.
Physical parameters, periods and period ratios for RRc stars are collected in Table~\ref{tab.rrc_periods}, sample of which is presented in Table~\ref{tab.rrc_periods_sample}. Parameters for RRd stars are collected in Table.~\ref{tab.rrd_periods}. We note that obtained physical parameters may not be representative for the whole population of RR~Lyrae stars. We studied stars in which one or two non-radial modes are detected. This specific mode selection may be possible for a specific range of physical parameters.

\begin{table*}
 \centering
  \caption{The results of the fitting of RRc stars. Consecutive columns provide the star's ID, physical parameters from the best-fitted model (mass, metallicity, effective temperature, luminosity), $D^2$ parameter, theoretical periods, and period ratios (top row), and observed periods and period ratios (bottom row). Full table is provided in the Appendix in Table~\ref{tab.rrc_periods} and in the machine readable format online.}~\\
  \label{tab.rrc_periods_sample}
  \centering
  \begin{tabular}{@{}lllllllll@{}}
\hline
\\
ID & $M$ [M$_\odot$] & [Fe/H] & log$T_{\rm eff}$ & log $L/{\rm L}_\odot$ & $D^2$ & $P_{\rm 1O}$ [d] & $R_8$ & $R_9$ \\ [0.2cm]
\hline
OGLE-BLG-RRLYR-06352 & 0.71 & -1.10 & 3.825 & 1.59 & 1.6$\cdot 10^{-7}$ & 0.3164 & 0.6311 & 0.6132\\ 
 & & & & & & 0.3163 & 0.6312 & 0.6132\\ 
OGLE-BLG-RRLYR-07803 & 0.68 & -0.75 & 3.830 & 1.59 & 2.1$\cdot 10^{-8}$ & 0.3139 & 0.6311 & 0.6123\\ 
 & & & & & & 0.3139 & 0.6311 & 0.6123\\ 
OGLE-BLG-RRLYR-08002 & 0.63 & -1.15 & 3.840 & 1.60 & 4.6$\cdot 10^{-8}$ & 0.3087 & 0.6307 & 0.6126\\ 
 & & & & & & 0.3088 & 0.6306 & 0.6126\\ 
OGLE-BLG-RRLYR-08475 & 0.64 & -1.30 & 3.840 & 1.61 & 6.0$\cdot 10^{-8}$ & 0.3110 & 0.6310 & 0.6132\\ 
 & & & & & & 0.3109 & 0.6311 & 0.6132\\ 
OGLE-BLG-RRLYR-08799 & 0.68 & -1.15 & 3.840 & 1.64 & 4.7$\cdot 10^{-8}$ & 0.3174 & 0.6311 & 0.6131\\ 
 & & & & & & 0.3173 & 0.6311 & 0.6130\\ 
 \multicolumn{1}{|c|}{\vdots}  &  \multicolumn{1}{|c|}{\vdots} & \multicolumn{1}{|c|}{\vdots} &  \multicolumn{1}{|c|}{\vdots} &  \multicolumn{1}{|c|}{\vdots} &  \multicolumn{1}{|c|}{\vdots} &  \multicolumn{1}{|c|}{\vdots} &  \multicolumn{1}{|c|}{\vdots} &  \multicolumn{1}{|c|}{\vdots}\\
 \hline
 \end{tabular}
 \end{table*}
 
 \begin{table*}
 \centering
  \caption{The results of the fitting of RRd stars. Consecutive columns provide star's ID, physical parameters from the best fitted model (mass, metallicity, effective temperature, luminosity), $D^2$ parameter, theoretical periods and period ratios (top row) and observed periods and period ratios (bottom row).}~\\
  \label{tab.rrd_periods}
  \centering
  \begin{tabular}{@{}lllllllllll@{}}
\hline
\\
ID & $M$ [M$_\odot$] & [Fe/H] & log$T_{\rm eff}$ & log $L/{\rm L_\odot}$ & $D^2$ & $P_{\rm F}$ [d] & $P_{\rm 1O}$ [d] & $R_8$ & $R_9$ \\ [0.2cm]
\hline
OGLE-BLG-RRLYR-09258	&	0.57	&	-0.40	&	3.805	&	1.36	&	1.8$\cdot 10^{-7}$	&	0.3793	&	0.2777	&	0.6308	&		\\
	&		&		&		&		&		&	0.3793	&	0.2776	&	0.6306	&		\\
OGLE-BLG-RRLYR-10369	&	0.79	&	-0.80	&	3.805	&	1.55	&	4.1$\cdot 10^{-7}$	&	0.4349	&	0.3217	&	0.6318	&		\\
	&		&		&		&		&		&	0.4350	&	0.3217	&	0.6321	&		\\
OGLE-BLG-RRLYR-13198	&	0.73	&	-1.05	&	3.840	&	1.67	&	1.4$\cdot 10^{-7}$	&	0.4344	&	0.3217	&	0.6316	&		\\
	&		&		&		&		&		&	0.4343	&	0.3217	&	0.6317	&		\\
OGLE-BLG-RRLYR-13666	&	0.68	&	-0.70	&	3.795	&	1.43	&	2.6$\cdot 10^{-7}$	&	0.4134	&	0.3050	&	0.6315	&		\\
	&		&		&		&		&		&	0.4132	&	0.3049	&	0.6315	&		\\
\hline																			
OGLE-BLG-RRLYR-10744	&	0.9	&	-1.10	&	3.815	&	1.69	&	2.5$\cdot 10^{-7}$	&	0.4781	&	0.3558	&		&	0.6142	\\
	&		&		&		&		&		&	0.4783	&	0.3559	&		&	0.6142	\\
OGLE-BLG-RRLYR-11234	&	0.89	&	-1.05	&	3.800	&	1.62	&	1.0$\cdot 10^{-7}$	&	0.4744	&	0.3530	&		&	0.6140	\\
	&		&		&		&		&		&	0.4745	&	0.3530	&		&	0.6142	\\
OGLE-BLG-RRLYR-14029	&	0.71	&	-0.85	&	3.815	&	1.55	&	2.7$\cdot 10^{-7}$	&	0.4299	&	0.3180	&		&	0.6127	\\
	&		&		&		&		&		&	0.4300	&	0.3181	&		&	0.6126	\\
CoRoT 0101368812	&	0.89	&	-1.25	&	3.825	&	1.74	&	4.6$\cdot 10^{-8}$	&	0.4880	&	0.3636	&		&	0.6141	\\
	&		&		&		&		&		&	0.4880	&	0.3637	&		&	0.6141	\\
M3 V13	&	0.85	&	-0.50	&	3.815	&	1.66	&	3.5$\cdot 10^{-7}$	&	0.4795	&	0.3507	&		&	0.6137	\\
	&		&		&		&		&		&	0.4794	&	0.3509	&		&	0.6137	\\
M3 V68	&	0.89	&	-1.20	&	3.840	&	1.79	&	2.1$\cdot 10^{-7}$	&	0.4785	&	0.3560	&		&	0.6145	\\
	&		&		&		&		&		&	0.4786	&	0.3559	&		&	0.6147	\\
\hline																			
OGLE-BLG-RRLYR-10796	&	0.69	&	-0.80	&	3.800	&	1.48	&	2.2$\cdot 10^{-7}$	&	0.4314	&	0.3191	&	0.6309	&	0.6124	\\
	&		&		&		&		&		&	0.4315	&	0.3190	&	0.6308	&	0.6122	\\
 \end{tabular}
 \end{table*}

\begin{figure}
     \centering
     \includegraphics[width=\columnwidth]{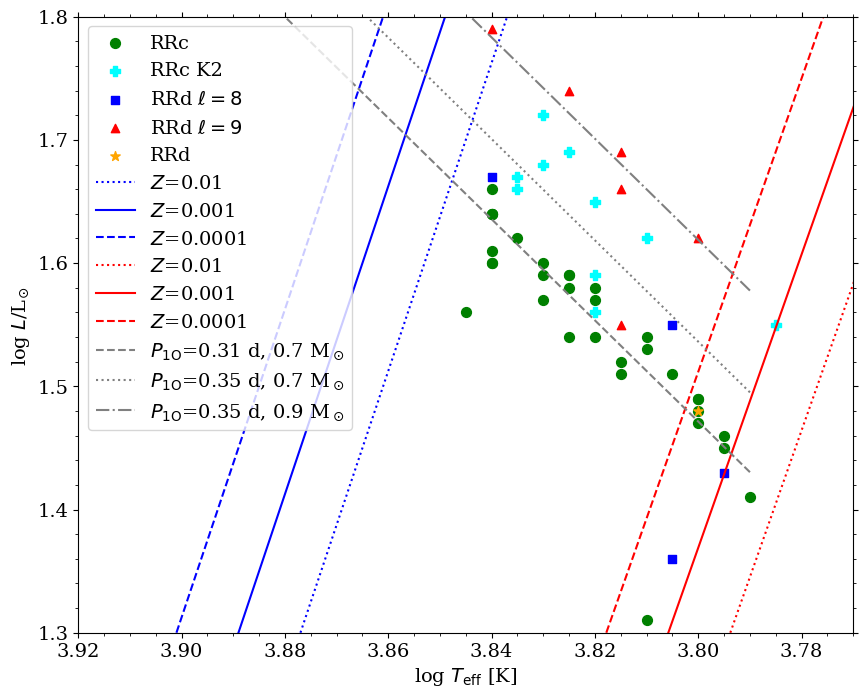}
     \caption{Hertzsprung-Russell diagrams for best-matching models to RRc and RRd stars with non-radial modes. RRc stars from the Galactic bulge and globular cluster NGC~6362 are plotted with green circles. RRc stars selected from the K2 data are plotted with cyan plus symbols. RRd stars with the non-radial mode of degree 8 are plotted with blue squares, and those with the non-radial mode of degree 9 are plotted with red triangles. The RRd star with both non-radial modes is plotted with an orange star. Blue and red edges for three values of metallicity are based on \protect\cite{marconi2015}.}
     \label{fig:hr_results}
 \end{figure}

 In Fig.~\ref{fig:hr_results}, we show our models in the HR diagram. Models cover the majority of the range in effective temperature of the instability strip, but they do not reach the highest effective temperatures. We did not find any models with an effective temperature above 7000\,K ($\log T_{\rm eff}>3.85$). This is consistent with the fact, that close to the blue edge of the instability strip there are no unstable non-radial modes (see Fig.~\ref{fig:grid_inst}). There are several models close to the red edge of the instability strip for the fundamental mode. We note that the red edges displayed in Fig.~\ref{fig:hr_results} correspond to three values of metallicity only. In addition, the position of the red edge is affected by the uncertainties of convection-pulsation coupling modeling. Qualitative comparison of positions of RR$_{0.61}$ stars with non-linear instability regions from \cite{szabo2004} does not lead to discrepancies for the majority of stars. It is also the case for similar comparison for RRc stars with $\log L/{\rm L}_\odot < 1.6$, not covered in the work by \cite{szabo2004} (Gabor Kovacs, private communication). RRd stars with the non-radial mode of degree 8 cover similar range of effective temperature and luminosity as RRc stars. RRd stars with the non-radial mode of degree 9, with an exception of two stars, have higher luminosity. They also have long periods of the first overtone. RRc stars from K2 sample also, on average, have slightly higher values of luminosity as compared to the RRc stars from the OGLE Galactic bulge sample. For the whole sample, we observe an increase in luminosity with increasing effective temperature. The slope is similar to the slope of lines of the constant period, as shown in Fig.~\ref{fig:hr_results}. The majority of RRc stars from the Galactic bulge have a period of the first overtone around 0.31\,d. Stars from the K2 mission have typically longer periods, around 0.34\,d, which is consistent with their higher luminosity. The majority of RRd stars with the non-radial mode of degree 9 have periods around 0.36\,d.

\begin{figure}
     \centering
     \includegraphics[width=\columnwidth]{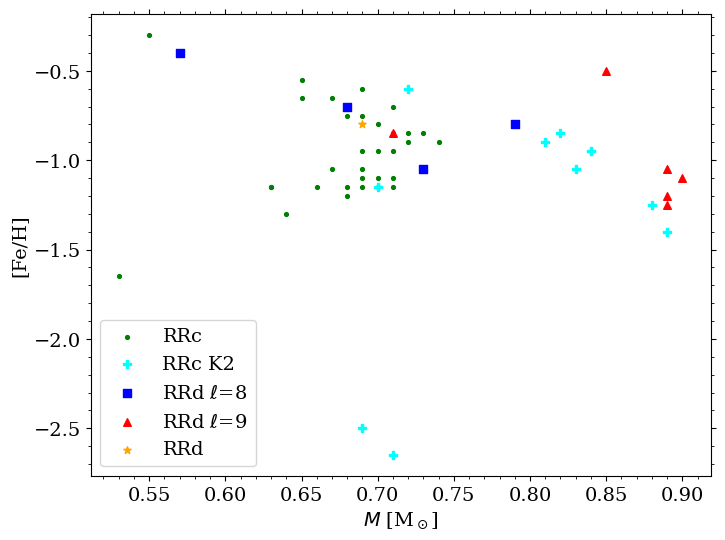}
     \caption{Metallicity--mass dependence for RRc and RRd models. The meaning of symbols is the same as in Fig.~\ref{fig:hr_results}.}
     \label{fig:mfeh}
 \end{figure}

 In Fig.~\ref{fig:mfeh}, we plotted the relationship between the metallicity and mass for the models for different groups of studied stars. The majority of the models have metallicities around $-1.0$. Only four stars, including three from the K2 sample, have metallicities below $-2.0$. There is no clear trend. The majority of stars have masses from 0.5\,M$_\odot$ to 0.75\,M$_\odot$, which is a range expected for RR~Lyrae stars. Higher masses, above 0.8\,M$_\odot$, were obtained for several stars from the K2 sample and for the majority of RRd stars with the non-radial mode of degree 9. Interestingly, RRd stars with the non-radial mode of degree 8 cover a similar range of parameters as RRc stars.
 
\begin{figure}
     \centering
     \includegraphics[width=\columnwidth]{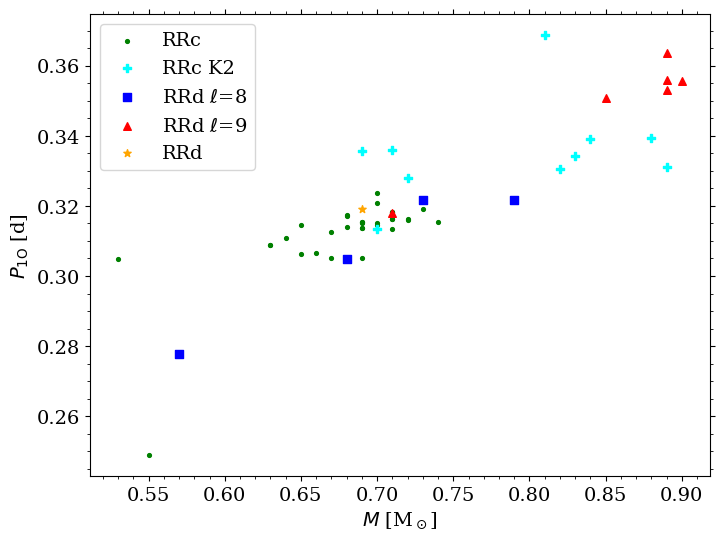}
     \caption{Relation between the observed first overtone periods and masses for RRc and RRd models with non-radial modes. The meaning of symbols is the same as in Fig.~\ref{fig:hr_results}.}
     \label{fig:mp1o}
 \end{figure}

 In Fig.~\ref{fig:mp1o}, we plotted the relationship between the observed period of the first overtone and masses of the models. Masses of the models increase with increasing observed period. Several RRc stars from the K2 sample and the majority of RRd stars with non-radial mode of degree 9, which have higher masses, have also long pulsation periods. 
 
\section{Discussion}\label{Sec.discussion}

\subsection{Fitting of the models}\label{Subsec.fitting}
Before  fitting the models to the observed sample, we compared the position of stars and of the grid of models on the Petersen diagram (see Fig.~\ref{fig:grid_pet}). Four stars were located between two sequences defined by the models. These stars are: OGLE-BLG-RRLYR-11981, EPIC~60018653, EPIC~60018662, and V87 from the globular cluster M3. Modeling of these stars was not possible. According to \cite{dziembowski2016}, the middle sequence on the Petersen diagram corresponds to the linear frequency combination between two non-radial modes, $f_8+f_9$. Based on the period ratios, these four stars were originally classified as belonging to the bottom sequence. However, comparison with the models suggests that these stars might belong to the middle sequence. In other words, in these four stars, we may observe only the combination frequency and not the harmonics of the non-radial modes. It is also possible, however, that these signals correspond to the harmonic of the non-radial mode, but the physical parameters are atypical and were not included in the model grid. Another (least likely) possibility is that the observed signals are false positive detections.

For further modeling, we used 45 RRc stars and 14 RRd stars. We found well-fitting models ($D^2<5\cdot 10^{-7}$) for 42 RRc and 11 RRd stars. Three RRc stars for which we did not find satisfactory models are OGLE-BLG-RRLYR-05202, OGLE-BLG-RRLYR-07665, and EPIC~212684145. In OGLE-BLG-RRLYR-05202 observed $R_8$ is higher than average value for RR$_{0.61}$ group (0.6356, where the average value is around 0.631), while $R_9$ is close to the average value (0.6127, where the average value is around 0.613). The reverse situation is observed for OGLE-BLG-RRLYR-07665. The lack of well-fitted models for these two stars may be a result of the incorrect classification of these stars as triple-mode, whereas they might be only double-mode. The third observed signal, corresponding to the atypical period ratio, may be a false positive or may arise due to contamination by nearby stars. Star from the K2 sample, EPIC~212684145, has typical values of period ratios as compared with the rest of the K2 sample. The best-fitting model has $D^2=7.1\cdot 10^{-7}$, so only slightly higher than the adopted threshold value of $D^2=5\cdot 10^{-7}$. In the case of EPIC~212684145 detailed modeling with a denser grid might result in a better model. Another possibility is that the frequencies of the additional signals in EPIC~212684145 might not be accurate due to the short time of observations (see later in this section).

We were not able to find well-fitted models for three RRd stars: OGLE-BLG-RRLYR-13721, OGLE-BLG-RRLYR-14031, and AQ~Leo. All three stars have atypical period ratios. Additionally, OGLE-BLG-RRLYR-14031 and AQ~Leo have the longest periods in the sample. RRd stars with such long periods are a challenge for modeling on their own \citep[see e.g.][]{smolec.prudil2016}. They would require high masses and low metallicity. From the initial sample selected for modeling, we were not able to model five stars with periods longer than 0.4~d, including stars that were rejected when compared to the model grid on the Petersen diagram (see Fig.~\ref{fig:grid_pet}).

Non-radial modes of degrees 8 and 9 and their harmonics are non-stationary signals. In frequency spectra, they manifest as groups of signals or power excess \citep[see e.g.][]{netzel_census}. It corresponds to the variable amplitude and frequency in the time domain. The variability may have different time scales depending on the star \citep{moskalik2015}. One of the reasons for the observed variability of non-radial modes are interactions between $2\ell+1$ components of the multiplet \citep{dziembowski2016}. When analyzing photometric data of a long time span, the wide groups of signals in frequency spectra are centered at the frequency corresponding to $m=0$. Calculated frequencies also correspond to $m=0$ because rotation and interaction within the multiplets are neglected. Therefore, fitting the Gaussian function to the structures in frequency spectra allows for estimation of the representative frequency, that can be used for comparison with the models. Estimation of the frequency in the case of photometric data of a shorter time span is not as reliable. K2 data covers on average 80 days of observations for each star. Hence, the results for the K2 sample of RRc stars can be affected by this effect.

In the literature, the frequency corresponding to the additional signal is usually the frequency of the highest signal from the structure observed in the frequency spectrum. This can affect the results for stars observed in the globular clusters M3 and NGC~6362.

  \begin{figure*}
     \centering
     \includegraphics[width=\textwidth]{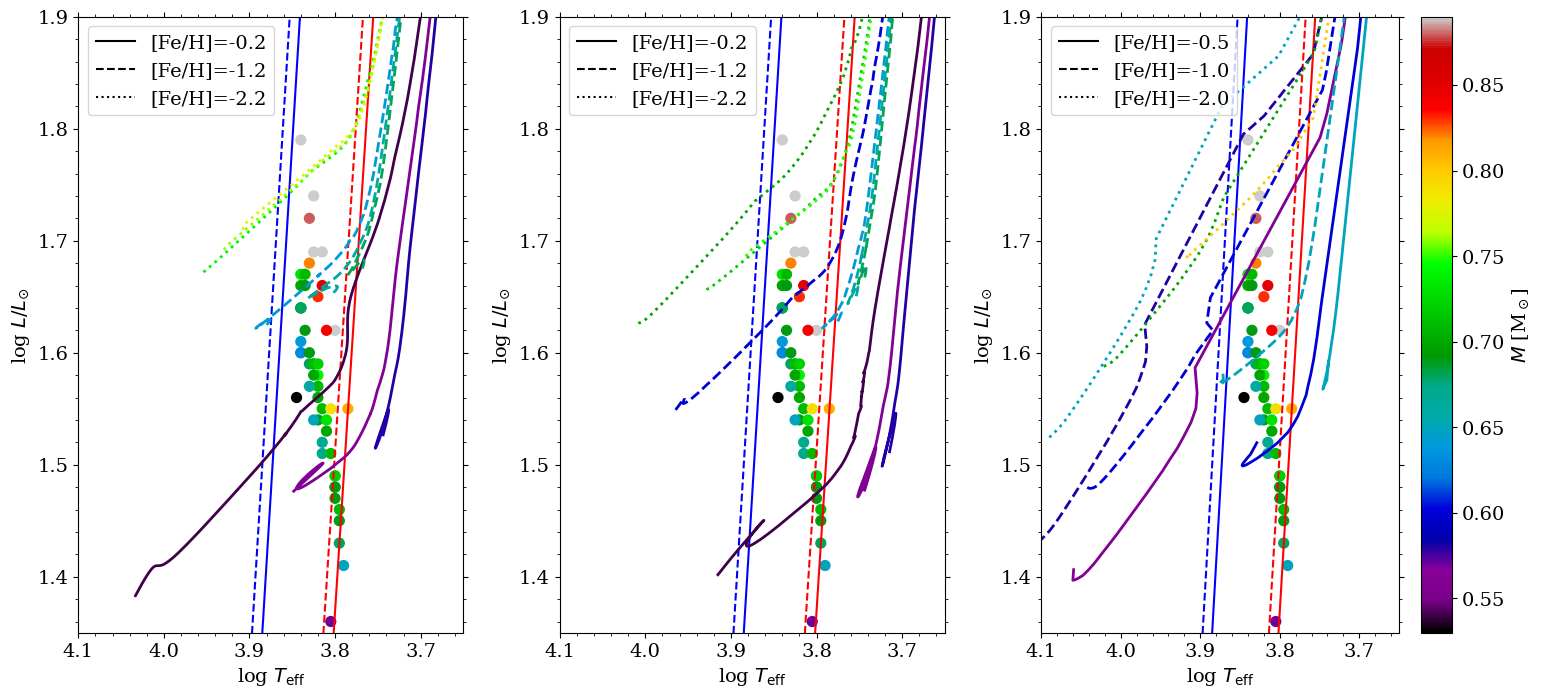}
     \caption{Comparison of models selected for RRc and RRd stars with non-radial modes, with evolutionary tracks on the Hertzsprung-Russell diagram. Masses of the models and evolutionary tracks are color-coded as indicated in the legend. Line type corresponds to the metallicity of evolutionary tracks. Left panel: evolutionary tracks from the BaSTI database \protect\citep{basti}. Middle panel: $\alpha$-enhanced evolutionary tracks from the BaSTI database. Right panel: evolutionary tracks from the Dartmouth database \protect\citep{dartmouth}. Blue and red edges of the instability strip are the same as in Fig.~\ref{fig:hr_results}.}
     \label{fig:comp_hr}
 \end{figure*}

\subsection{Comparison with evolution theory}

 Modeling the evolution of RR~Lyrae stars is not a trivial task, difficulties including uncertainties related to modeling of helium flash or mass loss on the red giant branch. In the pulsation modeling, we used an envelope code, which means that the physical parameters of the models are not connected to the evolution theory. Therefore, it is interesting to compare the results of the modeling with the prediction of the evolution theory. Such comparison in the case of classical Cepheids leads to the mass discrepancy problem \citep[e.g.][]{cox1980}. Masses predicted by the evolution theory are around 10-20\% higher than those predicted by the pulsation theory \citep[e.g.,][]{keller2008}.  In the case of RR Lyrae stars, \cite{kovacs2021} combined pulsation modeling with evolutionary tracks for single RR Lyrae stars to derive luminosities, which were later compared with estimations based on the Gaia EDR3 data, and led to an agreement between the two methods.
 
\begin{figure}
     \centering
     \includegraphics[width=\columnwidth]{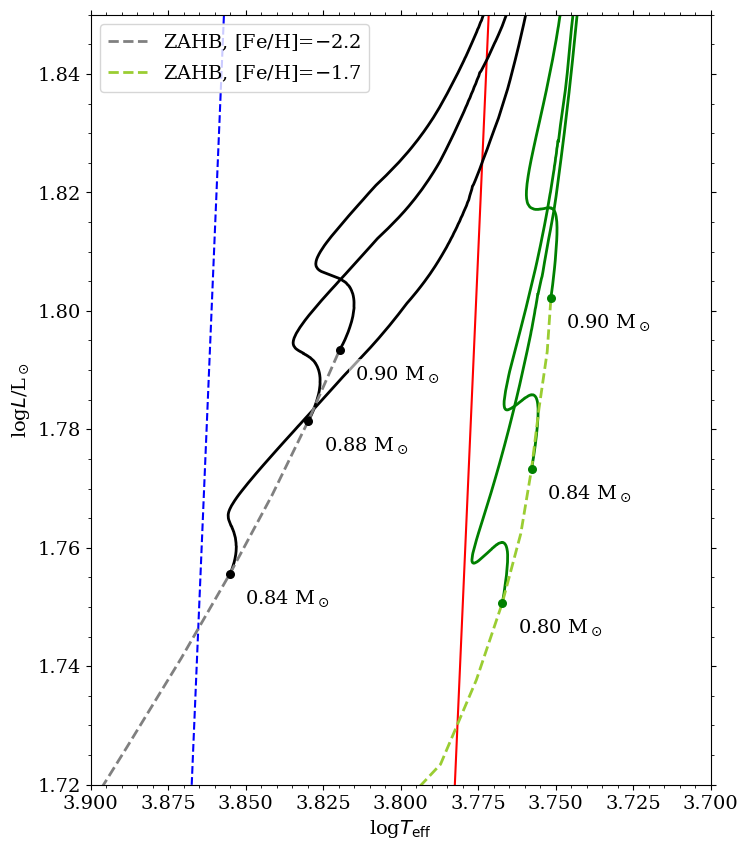}
     \caption{Evolutionary tracks for massive horizontal-branch stars based on the BaSTI database \protect\citep{basti}. Dashed green and black lines correspond to zero-age horizontal branch (ZAHB) for two different values of metallicity as indicated in the key. Blue and red edges of the instability strip are the same as in Fig.~\ref{fig:hr_results}.}
     \label{fig:rr_is}
 \end{figure}

 Comparison of the results obtained in the pulsation modeling with evolutionary tracks is presented in the HR diagram in Fig.~\ref{fig:comp_hr}. We used horizontal branch evolutionary tracks from BaSTI \citep{basti} with and without $\alpha$-enhancement, and Dartmouth databases \citep{dartmouth}. Qualitative comparison of the position of models with the evolutionary tracks of similar metallicity suggests that masses determined in the pulsation modeling are higher than masses corresponding to the evolutionary tracks located in the same regions of the instability strip. The highest discrepancy is observed for pulsation masses above 0.8\,M$_\odot$. The $\alpha$-enhanced tracks are shifted towards lower luminosities and lower temperatures, which seems to slightly reduce the observed discrepancy. Evolutionary tracks for masses above 0.8\,M$_\odot$ are shown in Fig.~\ref{fig:rr_is}. For such high masses, only low-metallicity tracks are within the edges of the instability strip. In the case of studied stars with higher masses derived in pulsation modeling, the mean metallicity is higher, around $-1.0$. On the other hand, \cite{kovacs1992} studied the effect of chemical composition on pulsation periods in RRd stars using pulsation models and reproduced two representative period ratios of Oosterhoff I and Oosterhoff II variables with masses above 0.8\,M$_\odot$ for metal abundance $Z=0.001$ ([Fe/H]$\approx$--1.13).

The reasons for obtaining such high masses for several stars from the K2 sample and for the majority of RRd stars with the non-radial mode of degree 9 are difficult to explain. We may speculate that in the case of stars with long pulsation periods, especially for RRd stars with the non-radial mode of degree 9, non-linear effects may affect the observed periods of non-radial modes. Note, that currently, we do not know any RRab star that shows non-radial modes of degrees 8 or 9. These modes are observed solely in stars with the first overtone excited (RRc or RRd). Linear pulsation theory does not explain why these modes are not observed in RRab stars. This suggests that fundamental mode can have a strong impact on the properties of considered non-radial modes. Unfortunately, we have no tools to study nonlinear mode interactions for radial and non-radial modes.

In the case of stars from the K2 sample, the observed frequencies might be affected by the short time span of photometric observations. Longer photometric observations could test this hypothesis (see the discussion at the end of Sec.~\ref{Subsec.fitting}).

The only observational differences between stars for which we obtained masses below and above 0.8\,M$_\odot$ are longer periods of pulsations, and slightly higher period ratios. Further investigation of this puzzling results would require more detailed modeling, preferably with additional constraints on the models, e.g. independent metallicity determination.

There is no direct mass determination for any known RR~Lyrae star, as no RR~Lyrae star is known to be in the eclipsing binary system. The most promising candidate for an RR~Lyrae star in a binary system turned out to be a star with significantly smaller mass and formed through a different evolutionary channel \citep{pietrzynski2012}. Search for RR~Lyrae stars in binary systems is ongoing and resulted in several candidates \citep[see e.g.][and references therein]{hajdu2018}. Unfortunately, binary systems detected using the light time effect will not yield dynamical masses for RR Lyrae stars. Without direct mass determination for RR Lyrae stars, we may further compare our mass estimates with masses determined indirectly, e.g. based on the shape of the light curve \citep{simon.clement1993}, on comparison with evolutionary tracks \citep[see e.g.][]{marsakov2019} or based on asteroseismic modeling \citep{molnar2015}.

\cite{simon.clement1993} inferred the relations between the physical parameters and the light curve shapes based on the hydrodynamic modeling of RRc stars in globular clusters. In the case of stars studied here, applying these relations leads to very low masses, below 0.45\,M$_\odot$, which is below the expected mass for RR~Lyrae stars. The work on connecting the physical parameters to light curve shapes using non-linear pulsation codes is ongoing \citep[see e.g.][]{bellinger2020}.

\cite{marsakov2019} estimated masses of nearly 100 RRc stars in globular clusters based on comparison with evolutionary tracks. Estimated masses are significantly lower than of the RR~Lyrae stars studied here. The average mass in the sample studied by \cite{marsakov2019} is 0.55\,M$_\odot$, whereas here we obtained the average mass of 0.71\,M$_\odot$. This is consistent with the already noticed discrepancy between masses predicted by the pulsation and evolution theory (see Fig.~\ref{fig:comp_hr}).

  \begin{figure}
     \centering
     \includegraphics[width=\columnwidth]{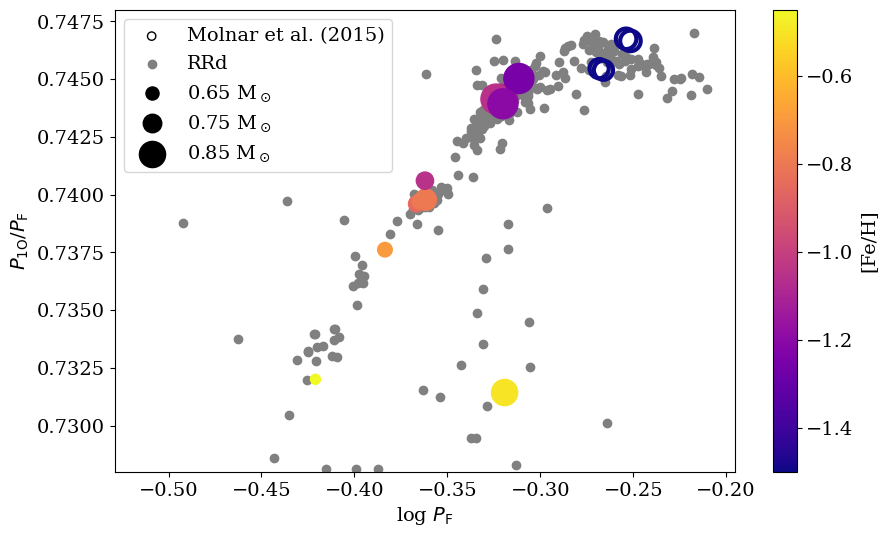}
     \caption{Petersen diagram for RRd stars. RRd stars and anomalous RRd stars from the Galactic bulge are plotted with gray points \protect\citep{soszynski2019}. RRd stars studied in this work are plotted with filled circles, where color corresponds to metallicity and size to mass. Two RRd stars studied by \protect\cite{molnar2015} are plotted with open circles with size and color plotted accordingly.}
     \label{fig:rrd_pet}
 \end{figure}

\begin{figure}
     \centering
     \includegraphics[width=\columnwidth]{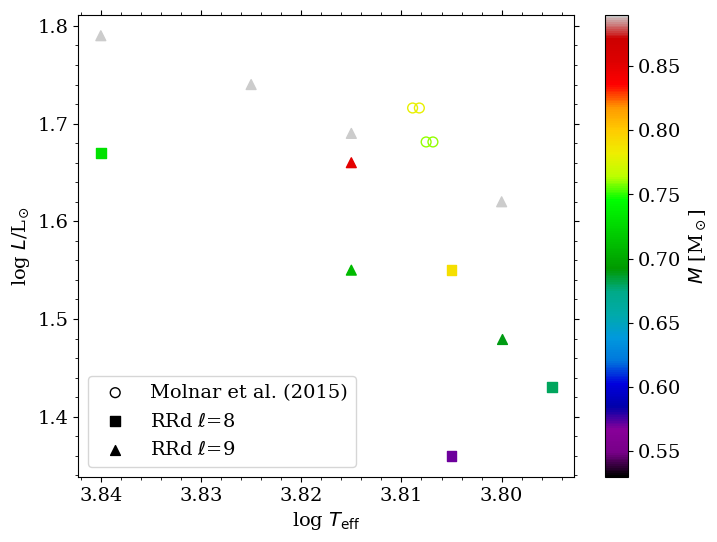}
     \caption{Hertzsprung-Russell diagram for RRd stars studied here and the two RRd stars studied by \protect\cite{molnar2015} (two models for each of two stars). Color of the symbols correspond to the mass obtained in the modeling, as indicated on the right side of the plot.}
     \label{fig:rrd_hr}
 \end{figure}

There are also mass estimations for individual RR~Lyrae stars based on asteroseismic modeling. \cite{molnar2015} used non-linear hydrodynamic modeling for two RRd stars and obtained masses of 0.76\,M$_\odot$ and 0.78\,M$_\odot$. Petersen diagram for radial modes in RRd stars from our sample and the study by \cite{molnar2015} is presented in Fig.~\ref{fig:rrd_pet}. We included information about the metallicity and mass as indicated in the key. For reference, we plotted all RRd stars from the Galactic bulge OGLE sample \citep{soszynski2019}. There is a decrease of period ratio with increasing metallicity and an increase of period ratio with increasing mass. Both these relations are expected for RRd stars based on the linear pulsation modeling (see e.g. \citealt{kovacs1992} and fig.~12 in \citealt{smolec.prudil2016}). The RRd star with an atypical period ratio is an anomalous RRd star V13 from the globular cluster M3. In Fig.~\ref{fig:rrd_hr} RRd stars analyzed here are plotted in the Hertzsprung-Russell diagram together with two RRd stars analyzed by \cite{molnar2015}. Effective temperature and luminosity for the two RRd stars are similar to those obtained for RRd stars with non-radial modes. We note however, that RRd stars analyzed by \cite{molnar2015} might be different than RRd stars analyzed here. On the Petersen diagram they are located on the part of the RRd sequence where the period ratio decreases with increasing fundamental mode period. All RRc stars analyzed in this study have shorter periods and are located in the part of the RRd sequence where we observe increasing period ratio with increasing fundamental mode period (see Fig.~\ref{fig:rrd_pet}).

 \subsection{Comparison with observations}
 
   \begin{figure}
     \centering
     \includegraphics[width=\columnwidth]{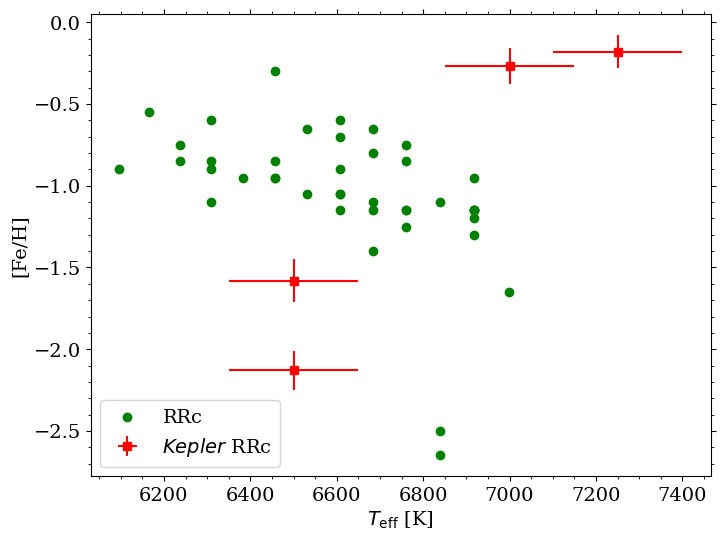}
     \caption{The ${\rm [Fe/H]}-T_{\rm eff}$ dependence for RRc stars with non-radial modes modeled in this work (green points) and four RRc stars from the original {\it Kepler} field. Parameters of these four stars were determined spectroscopically by \protect\cite{nemec2013} (red squares).}
     \label{fig:rrc_nemec}
 \end{figure}
 
Estimations of effective temperature and metallicity for RR~Lyrae stars, based on spectroscopic observations, are available. \cite{nemec2013} obtained these parameters for four RRc stars observed in the original {\it Kepler} field. The additional signals corresponding to the non-radial modes are observed in all four stars \citep{moskalik2015}. These stars were not included in the initial sample for the modeling, because only single additional signal corresponding to the RR$_{0.61}$ group was detected in each star. In Fig.~\ref{fig:rrc_nemec} metallicity and effective temperature of the four {\it Kepler} RRc stars is compared with parameters obtained in the modeling of RRc stars with non-radial modes. One star from the {\it Kepler} field has effective temperature higher than 7000\,K. The other three stars have effective temperatures from the range obtained in this modeling. Two {\it Kepler} stars have also relatively high metallicity (${\rm [Fe/H]}=-0.27$ and ${\rm [Fe/H]}=-0.18$), whereas the highest metallicity for the RRc sample with non-radial modes is ${\rm [Fe/H]}=-0.3$.
 
 \begin{figure}
     \centering
     \includegraphics[width=\columnwidth]{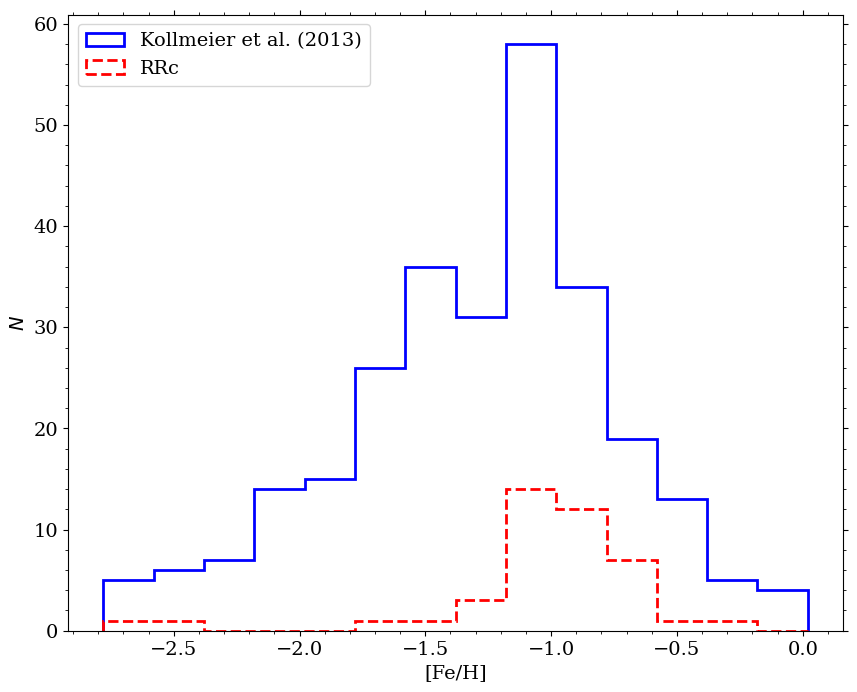}
     \caption{The distribution of metallicity for RRc stars. Blue solid line: metallicity determined in spectroscopic observations of RRc stars in the solar neighbourhood by \protect\cite{kollmeier2013}. Red dashed line: metallicity obtained in the asteroseismic modeling in this study.}
     \label{fig:feh_hist}
 \end{figure}
 
 \cite{kollmeier2013} obtained spectroscopic metallicity for nearly 300 RRc stars in the solar neighbourhood. In Fig.~\ref{fig:feh_hist} we plotted the distribution of the metallicity for RRc stars from the study by \cite{kollmeier2013} and for RRc stars with non-radial modes analyzed here. Both distributions have similar shape and a maximum around ${\rm [Fe/H]}=-1.1$.

 \subsection{The incidence rates of RR$_{0.61}$ stars}
 
 The incidence rate of RR$_{0.61}$ stars depends on the study: what system is analyzed and what data are used. When using a ground-based photometry, the incidence rate increases with increasing quality of the photometry. 
 The analysis of the RRc stars from the Galactic bulge based on the OGLE-III data resulted in the incidence rate of 2.9 per cent \citep{netzel1}. With OGLE-IV data, the incidence rate for the Galactic bulge is 8.7 per cent \citep{netzel_census}. \cite{netzel_2fields} studied only two best-sampled observing fields from OGLE-IV in the Galactic bulge which resulted in a significantly higher incidence rate of 27 per cent. 
 The analysis of the ground-based photometry for globular cluster M3 resulted in the incidence rate of RR$_{0.61}$ stars of 38 per cent \citep{jurcsik2015}. Globular cluster NGC~6362 was analyzed by \cite{smolec2017} using ground-based photometry and this study resulted in the incidence rate of 63 per cent. The incidence rates are even higher when analysing the excellent quality space-based photometry. The additional signal was found in all four RRc stars from the original {\it Kepler} field, giving the incidence rate of 100 per cent \citep{moskalik2015}. \cite{molnar2015} analyzed test K2 data for RR~Lyrae stars in a field in Pisces and obtained incidence rate of RR$_{0.61}$ stars of 75 per cent. First results on the RR~Lyrae stars observed in the TESS mission provided incidence rate of 65 per cent \citep{molnar2021}. General conclusion that can be drawn from these results is that the better the quality of the photometry, the higher the incidence rate. This raises a question whether in all first-overtone RR~Lyrae stars additional signals are excited, but sometimes cannot be detected due to noise level in frequency spectra. Color-magnitude diagrams of globular clusters M3 and NGC~6362 suggest that RR$_{0.61}$ tend to avoid the hottest (bluest) part of the diagrams \citep{jurcsik2015,smolec2017}. Similar trend was noted for the TESS RR$_{0.61}$ stars \citep{molnar2021}. Thus, observations indicate that RR$_{0.61}$ phenomenon, while widespread, is not ubiquitous. 
 
 Theoretical grid of models calculated in this work confirms the conclusions drawn from the observational color-magnitude diagrams. Indeed, close to the blue edge of the instability strip for the first overtone, non-radial modes are linearly stable (see Fig.~\ref{fig:grid_inst}). The lack of detection of non-radial modes in some of RR Lyrae stars might be intrinsic. Still, the part of the instability strip in which non-radial modes are unstable is large. Consequently, in the majority of first-overtone RR Lyrae stars the RR$_{0.61}$ type of pulsation is possible, which is consistent with high incidence rate of RR$_{0.61}$ stars in the space-based photometric data.

 \section{Conclusions}\label{Sec.conclusions}

 We performed asteroseismic modeling of RR$_{0.61}$ stars, which are RR Lyrae stars in which the first overtone is present together with the additional signals that form period ratio around 0.61 with the first-overtone period. We assumed that those signals are due to harmonics of non-radial modes of degrees 8 or 9 as proposed by \cite{dziembowski2016}. For the modeling we chose at least triple-mode stars. These can be either RRc stars with both non-radial modes or RRd stars with at least one additional non-radial mode. The input sample consists of 45 RRc and 18 RRd stars. 
 We calculated the grid of pulsation models for RR~Lyrae stars using the Warsaw envelope code \citep{dziembowski1977}. Then, we selected the models that reproduce observed periods and period ratios best. Physical parameters that were obtained are: effective temperature, luminosity, mass and metallicity. The results are following:
 \begin{itemize}
 \item We were able to fit models to 42 RRc and 11 RRd stars with additional non-radial modes. 
 \item  Four RRd stars were excluded from the modeling, because values of their period ratios were not covered by the grid of models. We were not able to find well fitted models for further 3 RRc and 3 RRd stars from the sample. Two RRc stars, for which we did not find matching models, have at least one atypical period ratio. Three RRd stars without well fitted models have period of the first overtone longer than 0.45\,d.
 \item The masses determined for RRc stars in the Galactic bulge are in the $0.53-0.74\,{\rm M}_\odot$ range. The masses for two stars observed in the globular cluster NGC~6362 are 0.66\,M$_\odot$ and 0.65\,M$_\odot$. RRc stars observed during the K2 mission have, on average, higher masses, from the $0.69-0.89\,{\rm M}_\odot$ range.
 \item The masses determined for RRd stars with additional non-radial mode of degree 8 are from the $0.57-0.79\,{\rm M}_\odot$ range. The masses for RRd stars with the additional non-radial mode of degree 9 are higher, from the $0.71-0.90\,{\rm  M}_\odot$ range. The only RRd star in which both non-radial modes were detected has a mass of 0.69\,M$_\odot$.
 \item We compared the positions of modeled RR~Lyrae stars with the evolutionary tracks for horizontal branch stars from BaSTI and Dartmouth databases on the Hertzsprung-Russell diagram. Qualitative comparison suggests that there is a mass discrepancy problem. For similar metallicities, masses estimated in the pulsation modeling are typically higher than masses predicted by the evolution theory.
 \item The strongest discrepancy is observed for longest period, most massive stars, with masses above 0.8\,M$_\odot$, for which evolutionary tracks do not fall within the instability strip (unless the metallicity is very low). These are mostly stars from K2 sample and RRd stars with non-radial mode of degree 9. Interestingly, RRd stars with the non-radial mode of degree 8, or one RRd stars which has both non-radial modes detected, have smaller masses, similar to the majority of RRc stars. The reasons for higher masses in some of the modelled stars are difficult to explain. It seems the only difference between stars for which we determine higher masses, and the rest of our sample, are the longer pulsation periods in the high-mass group. We  speculate that nonlinear effects in long period stars may result in frequency shifts of non-radial modes, and hence may limit the possibility of modelling long-period RR$_{0.61}$ stars at the linear level.  More photometric observations and larger sample for modeling, together with additional, independent constraints on some physical parameters, e.g. on metallicity, are necessary to study this problem in more detail. Independent mass determinations for RR~Lyrae stars would be extremely valuable to shed more light on the puzzle, as evolutionary masses are also subject to uncertainties.
  \item  We compared the distribution of metallicity for the  modeled sample with the distribution of metallicity based on spectroscopic measurements for RRc stars in the solar neighbourhood \citep{kollmeier2013}. Both distribution have similar shapes and have a maximum around ${\rm [Fe/H]}=-1.1$.
 \item Theoretical pulsation models show that there are no unstable non-radial modes of degrees 8 and 9 very close to the blue edge of the instability strip for the radial first overtone. This supports the conclusions based on color-magnitude diagrams for globular clusters \citep{jurcsik2015,smolec2017}: RR$_{0.61}$ are not detected on the blue side of the diagrams. The excitation of non-radial modes in RRc stars is very common, but not ubiquitous.
  \end{itemize}

The presented study is the very first attempt of asteroseismic modeling of RR~Lyrae stars using additional non-radial modes assuming that their identification by \cite{dziembowski2016} is correct. In our modelling we used a grid approach which allows us to study the statistical properties of the whole sample, in particular the distributions of physical parameters of the modelled stars. Our study should not be considered as detailed and precise modeling of individual objects.  We consider this first attempt successful. For the majority of stars the physical parameters are reasonable, within a range expected for RR~Lyrae stars. Comparison of the determined masses with those predicted by evolutionary tracks revealed a mass discrepancy problem, which should be quantified and studied in more detail. That should include detailed studies on the effect of $\alpha$-enhancement.  Our results are a good base for further study and for precise asteroseismic modeling of individual stars, in particular, when additional observational constraints are available.

\section*{Acknowledgements}

This project has been supported by the Lend\"ulet Program of the Hungarian Academy of Sciences, project No. LP2018-7/2020. HN is supported in by the Polish Ministry of Science and Higher Education under grant 0192/DIA/2016/45 within the Diamond Grant Program for years 2016-2021 and Foundation for Polish Science (FNP). RS was supported by the National Science Center, Poland, Sonata BIS project 2018/30/E/ST9/00598. This work has made use of BaSTI web tools. We thank Gabor Kovacs for fruitful discussions.

\section*{Data Availability}

Photometric data used for this study are published and references are provided. The grid of pulsation models will be shared upon a reasonable request to the corresponding author.



\bibliographystyle{mnras}
\bibliography{references} 




\appendix

\section{Physical parameters}

\begin{table*}
 \centering
  \caption{The results of the fitting of RRc stars. Consecutive columns provide star's ID, physical parameters from the best fitted model (mass, metallicity, effective temperature, luminosity), $D^2$ parameter, theoretical periods and period ratios (top row) and observed periods and period ratios (bottom row).}~\\
  \label{tab.rrc_periods}
  \centering
  \begin{tabular}{@{}lllllllll@{}}
\hline
\\
ID & $M$ [M$_\odot$] & [Fe/H] & log$T_{\rm eff}$ & log $\frac{L}{\rm L_\odot}$ & $D^2$ & $P_{\rm 1O}$ [d] & $R_8$ & $R_9$ \\ [0.2cm]
\hline
OGLE-BLG-RRLYR-06352 & 0.71 & -1.10 & 3.825 & 1.59 & 1.6$\cdot 10^{-7}$ & 0.3164 & 0.6311 & 0.6132\\ 
 & & & & & & 0.3163 & 0.6312 & 0.6132\\ 
OGLE-BLG-RRLYR-07803 & 0.68 & -0.75 & 3.830 & 1.59 & 2.1$\cdot 10^{-8}$ & 0.3139 & 0.6311 & 0.6123\\ 
 & & & & & & 0.3139 & 0.6311 & 0.6123\\ 
OGLE-BLG-RRLYR-08002 & 0.63 & -1.15 & 3.840 & 1.60 & 4.6$\cdot 10^{-8}$ & 0.3087 & 0.6307 & 0.6126\\ 
 & & & & & & 0.3088 & 0.6306 & 0.6126\\ 
OGLE-BLG-RRLYR-08475 & 0.64 & -1.30 & 3.840 & 1.61 & 6.0$\cdot 10^{-8}$ & 0.3110 & 0.6310 & 0.6132\\ 
 & & & & & & 0.3109 & 0.6311 & 0.6132\\ 
OGLE-BLG-RRLYR-08799 & 0.68 & -1.15 & 3.840 & 1.64 & 4.7$\cdot 10^{-8}$ & 0.3174 & 0.6311 & 0.6131\\ 
 & & & & & & 0.3173 & 0.6311 & 0.6130\\ 
OGLE-BLG-RRLYR-08920 & 0.70 & -1.10 & 3.800 & 1.48 & 6.6$\cdot 10^{-8}$ & 0.3150 & 0.6310 & 0.6132\\ 
 & & & & & & 0.3150 & 0.6311 & 0.6132\\ 
OGLE-BLG-RRLYR-09733 & 0.69 & -0.60 & 3.800 & 1.47 & 4.0$\cdot 10^{-8}$ & 0.3138 & 0.6313 & 0.6123\\ 
 & & & & & & 0.3138 & 0.6313 & 0.6124\\ 
OGLE-BLG-RRLYR-11072 & 0.72 & -0.90 & 3.800 & 1.49 & 7.5$\cdot 10^{-8}$ & 0.3161 & 0.6313 & 0.6131\\ 
 & & & & & & 0.3160 & 0.6312 & 0.6131\\ 
OGLE-BLG-RRLYR-11728 & 0.69 & -1.10 & 3.835 & 1.62 & 6.9$\cdot 10^{-8}$ & 0.3152 & 0.6311 & 0.6132\\ 
 & & & & & & 0.3152 & 0.6312 & 0.6133\\ 
OGLE-BLG-RRLYR-12261 & 0.69 & -1.05 & 3.820 & 1.54 & 9.1$\cdot 10^{-8}$ & 0.3051 & 0.6312 & 0.6132\\ 
 & & & & & & 0.3051 & 0.6310 & 0.6131\\ 
OGLE-BLG-RRLYR-31736 & 0.67 & -1.05 & 3.815 & 1.51 & 9.9$\cdot 10^{-8}$ & 0.3053 & 0.6308 & 0.6128\\ 
 & & & & & & 0.3052 & 0.6309 & 0.6130\\ 
OGLE-BLG-RRLYR-06802 & 0.71 & -0.70 & 3.820 & 1.57 & 3.7$\cdot 10^{-8}$ & 0.3183 & 0.6313 & 0.6126\\ 
 & & & & & & 0.3183 & 0.6314 & 0.6126\\ 
OGLE-BLG-RRLYR-06922 & 0.63 & -1.15 & 3.840 & 1.60 & 5.2$\cdot 10^{-8}$ & 0.3087 & 0.6307 & 0.6126\\ 
 & & & & & & 0.3088 & 0.6308 & 0.6126\\ 
OGLE-BLG-RRLYR-07047 & 0.53 & -1.65 & 3.845 & 1.56 & 6.3$\cdot 10^{-8}$ & 0.3050 & 0.6301 & 0.6124\\ 
 & & & & & & 0.3050 & 0.6302 & 0.6123\\ 
OGLE-BLG-RRLYR-07806 & 0.73 & -0.85 & 3.810 & 1.54 & 9.7$\cdot 10^{-8}$ & 0.3189 & 0.6313 & 0.6130\\ 
 & & & & & & 0.3190 & 0.6312 & 0.6131\\ 
OGLE-BLG-RRLYR-08826 & 0.67 & -0.65 & 3.815 & 1.52 & 1.4$\cdot 10^{-7}$ & 0.3125 & 0.6308 & 0.6118\\ 
 & & & & & & 0.3125 & 0.6309 & 0.6120\\ 
OGLE-BLG-RRLYR-08980 & 0.69 & -0.75 & 3.795 & 1.45 & 5.9$\cdot 10^{-9}$ & 0.3135 & 0.6312 & 0.6126\\ 
 & & & & & & 0.3136 & 0.6312 & 0.6126\\ 
OGLE-BLG-RRLYR-09444 & 0.70 & -1.15 & 3.840 & 1.66 & 10.0$\cdot 10^{-8}$ & 0.3237 & 0.6312 & 0.6133\\ 
 & & & & & & 0.3238 & 0.6312 & 0.6133\\ 
OGLE-BLG-RRLYR-10119 & 0.69 & -0.95 & 3.840 & 1.64 & 7.6$\cdot 10^{-8}$ & 0.3154 & 0.6312 & 0.6130\\ 
 & & & & & & 0.3154 & 0.6311 & 0.6130\\ 
OGLE-BLG-RRLYR-10262 & 0.71 & -1.15 & 3.825 & 1.59 & 1.5$\cdot 10^{-7}$ & 0.3162 & 0.6310 & 0.6132\\ 
 & & & & & & 0.3162 & 0.6311 & 0.6131\\ 
OGLE-BLG-RRLYR-10534 & 0.74 & -0.90 & 3.820 & 1.58 & 8.1$\cdot 10^{-8}$ & 0.3154 & 0.6314 & 0.6133\\ 
 & & & & & & 0.3153 & 0.6315 & 0.6132\\ 
OGLE-BLG-RRLYR-10614 & 0.70 & -0.80 & 3.825 & 1.58 & 1.5$\cdot 10^{-7}$ & 0.3144 & 0.6313 & 0.6127\\ 
 & & & & & & 0.3145 & 0.6311 & 0.6127\\ 
OGLE-BLG-RRLYR-11547 & 0.65 & -0.65 & 3.825 & 1.54 & 1.9$\cdot 10^{-7}$ & 0.3061 & 0.6307 & 0.6117\\ 
 & & & & & & 0.3062 & 0.6307 & 0.6115\\ 
OGLE-BLG-RRLYR-11621 & 0.68 & -1.20 & 3.840 & 1.64 & 9.7$\cdot 10^{-8}$ & 0.3172 & 0.6311 & 0.6132\\ 
 & & & & & & 0.3172 & 0.6311 & 0.6131\\ 
OGLE-BLG-RRLYR-11913 & 0.71 & -0.95 & 3.805 & 1.51 & 1.3$\cdot 10^{-7}$ & 0.3183 & 0.6310 & 0.6128\\ 
 & & & & & & 0.3184 & 0.6312 & 0.6130\\ 
OGLE-BLG-RRLYR-12769 & 0.71 & -0.85 & 3.795 & 1.46 & 9.0$\cdot 10^{-8}$ & 0.3134 & 0.6315 & 0.6132\\ 
 & & & & & & 0.3134 & 0.6313 & 0.6132\\ 
OGLE-BLG-RRLYR-12776 & 0.69 & -1.15 & 3.830 & 1.60 & 8.2$\cdot 10^{-8}$ & 0.3154 & 0.6310 & 0.6133\\ 
 & & & & & & 0.3154 & 0.6310 & 0.6132\\ 
OGLE-BLG-RRLYR-12972 & 0.70 & -0.95 & 3.810 & 1.53 & 6.2$\cdot 10^{-8}$ & 0.3209 & 0.6307 & 0.6125\\ 
 & & & & & & 0.3209 & 0.6308 & 0.6125\\ 
\hline
\end{tabular}
\end{table*}

\begin{table*}
 \centering
  \begin{tabular}{@{}lllllllll@{}}
\hline
\\
ID & $M$ [M$_\odot$] & [Fe/H] & log$T_{\rm eff}$ & log $\frac{L}{\rm L_\odot}$ & $D^2$ & $P_{\rm 1O}$ [d] & $R_8$ & $R_9$ \\ [0.2cm]
\hline
OGLE-BLG-RRLYR-13156 & 0.72 & -0.85 & 3.800 & 1.49 & 2.3$\cdot 10^{-8}$ & 0.3163 & 0.6312 & 0.6129\\ 
 & & & & & & 0.3163 & 0.6313 & 0.6129\\ 
OGLE-BLG-RRLYR-32213 & 0.55 & -0.30 & 3.810 & 1.31 & 4.4$\cdot 10^{-8}$ & 0.2489 & 0.6323 & 0.6126\\ 
 & & & & & & 0.2489 & 0.6323 & 0.6127\\ 
EPIC 210438688 & 0.72 & -0.60 & 3.820 & 1.59 & 8.0$\cdot 10^{-9}$ & 0.3280 & 0.6311 & 0.6121\\ 
 & & & & & & 0.3280 & 0.6311 & 0.6121\\ 
EPIC 212316775 & 0.82 & -0.85 & 3.830 & 1.68 & 8.1$\cdot 10^{-8}$ & 0.3305 & 0.6324 & 0.6141\\ 
 & & & & & & 0.3305 & 0.6323 & 0.6142\\ 
EPIC 211701322 & 0.84 & -0.95 & 3.810 & 1.62 & 1.5$\cdot 10^{-7}$ & 0.3392 & 0.6319 & 0.6139\\ 
 & & & & & & 0.3391 & 0.6317 & 0.6140\\ 
EPIC 211728918 & 0.88 & -1.25 & 3.830 & 1.72 & 5.0$\cdot 10^{-8}$ & 0.3395 & 0.6320 & 0.6147\\ 
 & & & & & & 0.3395 & 0.6321 & 0.6148\\ 
EPIC 212347262 & 0.83 & -1.05 & 3.820 & 1.65 & 1.3$\cdot 10^{-7}$ & 0.3343 & 0.6319 & 0.6141\\ 
 & & & & & & 0.3342 & 0.6320 & 0.6143\\ 
EPIC 212352472 & 0.71 & -2.65 & 3.835 & 1.67 & 2.5$\cdot 10^{-7}$ & 0.3359 & 0.6317 & 0.6155\\ 
 & & & & & & 0.3359 & 0.6320 & 0.6156\\ 
EPIC 212419731 & 0.69 & -2.50 & 3.835 & 1.66 & 2.8$\cdot 10^{-8}$ & 0.3356 & 0.6318 & 0.6153\\ 
 & & & & & & 0.3356 & 0.6318 & 0.6152\\ 
EPIC 212448152 & 0.89 & -1.40 & 3.825 & 1.69 & 4.0$\cdot 10^{-8}$ & 0.3311 & 0.6316 & 0.6149\\ 
 & & & & & & 0.3310 & 0.6317 & 0.6150\\ 
EPIC 212613425 & 0.81 & -0.90 & 3.785 & 1.55 & 7.6$\cdot 10^{-8}$ & 0.3688 & 0.6310 & 0.6127\\ 
 & & & & & & 0.3687 & 0.6310 & 0.6128\\ 
EPIC 212824246 & 0.70 & -1.15 & 3.820 & 1.56 & 3.4$\cdot 10^{-8}$ & 0.3134 & 0.6310 & 0.6132\\ 
 & & & & & & 0.3134 & 0.6309 & 0.6132\\ 
NGC 6362 V17 & 0.65 & -0.55 & 3.790 & 1.41 & 3.1$\cdot 10^{-7}$ & 0.3146 & 0.6306 & 0.6114\\ 
 & & & & & & 0.3146 & 0.6309 & 0.6117\\ 
NGC 6362 V33 & 0.66 & -1.15 & 3.830 & 1.57 & 3.7$\cdot 10^{-8}$ & 0.3065 & 0.6309 & 0.6130\\ 
 & & & & & & 0.3064 & 0.6309 & 0.6130\\ 
\hline
\end{tabular}
\end{table*}


\bsp	
\label{lastpage}
\end{document}